\documentclass[11pt,letterpaper]{article}
\usepackage{jheppub}
\usepackage[english]{babel}
\usepackage{amssymb,amsmath}
\usepackage{mathtools}
\usepackage{mathrsfs}
\usepackage{bbold}
\usepackage{hyperref}
\usepackage{empheq}
\usepackage{subfigure}
\usepackage{xfrac} 
\usepackage{graphics,color}
\usepackage{slashed}
\usepackage{graphicx}
\usepackage{braket}
\usepackage{float}
\usepackage{tcolorbox}
\usepackage{booktabs}
\usepackage[normalem]{ulem}

\newcommand{\be}{\begin{eqnarray}}
\newcommand{\ee}{\end{eqnarray}}
\newcommand{\nn}{\nonumber}

\usepackage{tikz}
\usetikzlibrary{arrows,shapes}
\usetikzlibrary{trees}
\usetikzlibrary{matrix,arrows} 				
\usetikzlibrary{positioning}				
\usetikzlibrary{calc,through}				
\usetikzlibrary{decorations.pathreplacing}  
\usepackage{pgffor}							

\usetikzlibrary{decorations.pathmorphing}	
\usetikzlibrary{decorations.markings}
\tikzset{
    vector/.style={decorate, decoration={snake}, draw},
	provector/.style={decorate, decoration={snake,amplitude=2.5pt}, draw},
	antivector/.style={decorate, decoration={snake,amplitude=-2.5pt}, draw},
	graviton/.style={decorate, decoration={snake,amplitude=1.5pt}, draw},
    fermion/.style={draw=black, postaction={decorate},
        decoration={markings,mark=at position .55 with {\arrow[draw=black]{>}}}},
    fermionbar/.style={draw=black, postaction={decorate},
        decoration={markings,mark=at position .55 with {\arrow[draw=black]{<}}}},
    fermionnoarrow/.style={draw=black},
    gluon/.style={decorate, draw=black,
        decoration={coil,amplitude=4pt, segment length=5pt}},
    scalar/.style={dashed,draw=black, postaction={decorate},
        decoration={markings,mark=at position .55 with {\arrow[draw=black]{>}}}},
    scalarbar/.style={dashed,draw=black, postaction={decorate},
        decoration={markings,mark=at position .55 with {\arrow[draw=black]{<}}}},
    scalarnoarrow/.style={dashed,draw=black},
    electron/.style={draw=black, postaction={decorate},
        decoration={markings,mark=at position .55 with {\arrow[draw=black]{>}}}},
	bigvector/.style={decorate, decoration={snake,amplitude=4pt}, draw},
}

\title{On the Dynamical Origin of the $\eta'$ Potential and the Axion Mass}
\preprint{}

\author[a]{Csaba Cs\'aki,}
\author[b]{Raffaele Tito D'Agnolo,}
\author[c]{Rick S. Gupta,}
\author[d]{Eric Kuflik,}
\author[c]{Tuhin S. Roy,}
\author[a]{and Maximilian Ruhdorfer\,}

\emailAdd{csaki@cornell.edu}
\emailAdd{raffaele-tito.dagnolo@ipht.fr}
\emailAdd{rsgupta@theory.tifr.res.in}
\emailAdd{eric.kuflik@mail.huji.ac.il}
\emailAdd{tuhin@theory.tifr.res.in}
\emailAdd{m.ruhdorfer@cornell.edu}

\affiliation[a]{Laboratory for Elementary Particle Physics, Cornell University, Ithaca, NY 14853, USA}
\affiliation[b]{Universit\'e Paris-Saclay, CEA, CNRS, Institut de Physique Th\'eorique, 91191, Gif-sur-Yvette, France}
\affiliation[c]{Department of Theoretical Physics, Tata Institute of Fundamental Research, Homi Bhabha Rd,
Mumbai 400005, India} 
\affiliation[d]{Racah Institute of Physics, Hebrew University of Jerusalem, Jerusalem 91904, Israel}

\abstract{We investigate the dynamics responsible for generating the potential of the $\eta'$, the (would-be) Goldstone boson associated with the anomalous axial $U(1)$ symmetry of QCD. The standard lore posits that pure QCD dynamics generates a confining potential with a branched structure as a function of the $\theta$ angle, and that this same potential largely determines the properties of the $\eta'$ once fermions are included. Here we test this picture by examining a supersymmetric extension of QCD with a small amount of supersymmetry breaking generated via anomaly mediation. For pure $SU(N)$ QCD without flavors, we verify that there are $N$ branches generated by gaugino condensation. Once quarks are introduced, the flavor effects qualitatively change the strong dynamics of the pure theory. For $F$ flavors we find $|N-F|$ branches, whose dynamical origin is gaugino condensation in the unbroken subgroup for $F<N-1$, and in the dual gauge group for $F>N+1$. For the special cases of $F=N-1,N,N+1$ we find no branches and the entire potential is consistent with being a one-instanton effect. The number of branches is a simple consequence of the selection rules of an anomalous $U(1)_R$ symmetry. We find that the $\eta'$ mass does not vanish in the large $N$ limit for fixed $F/N$, since the anomaly is non-vanishing. The same dynamics that is responsible for the $\eta'$ potential is also responsible for the axion potential.  We present a simple derivation of the axion mass formula for an arbitrary number of flavors.}

\makeatletter
\@addtoreset{footnote}{section}
\makeatother

\begin{document}

\maketitle	
\flushbottom

\section{Introduction}\label{sec:intro}

Understanding the dynamics of QCD is one the most important and exciting open questions in modern particle physics. While many qualitative aspects are clear, including the existence of chiral symmetry breaking with various quark and glueball condensates, the details of the dynamics leading to confinement and chiral symmetry breaking are still not fully understood. Instantons might be expected to play an important role in the confining dynamics. However, Witten, and separately Di Vecchia and Veneziano, argued convincingly that they are likely not responsible for the main features of the QCD $\eta^\prime$ potential~\cite{Witten:1978bc, DiVecchia:1980yfw}. Witten's argument focuses on the $\theta$-dependence of the QCD vacuum energy. In the absence of fermions (pure QCD) the dynamics responsible for confinement is expected to produce a non-vanishing potential which depends on the $\theta$ angle. Once massless fermions are introduced, the $\theta$-dependence should disappear. This is due to the presence of the $\eta'$ (the Goldstone boson of the anomalous axial symmetry) which acts as a heavy axion canceling the $\theta$-dependence of the theory.  However in the large $N$ limit the anomaly vanishes, and the $\eta'$ is expected to become massless. This has very important consequences on the $\theta$-dependence of the theory, which we will review in detail in Section~\ref{sec:largeN}. These arguments involving the large $N$ limit and the $\eta'$ mass led Witten to conclude that the potential of large $N$ QCD must have a branched structure, also implying that the dynamics responsible for generating the QCD potential is not instantons (since an instanton effect produces a potential that is strictly $2\pi$ periodic in $\theta$, and does not have a branched structure). This also implies that any attempts at deriving the usual QCD axion mass formula using insertions of 't Hooft operators due to instanton effects are futile. 

The first aim of this paper is to review the arguments summarized above in the simplest language of the chiral Lagrangian, and examine the consequences for axion dynamics. The main original part of the paper is the examination of the statements regarding confining dynamics, the $\eta'$ potential and its branched structure in a QCD-like theory obtained by perturbing the supersymmetric (SUSY) extension of QCD via small SUSY breaking obtained from Anomaly Mediated Supersymmetry Breaking (AMSB)~\cite{Randall:1998uk,Giudice:1998xp,Arkani-Hamed:1998mzz,Luty:1999qc,Pomarol:1999ie}. 

The dynamics of the unbroken SUSY theories have been understood in Ref.~\cite{Affleck:1983mk,Seiberg:1994bz,Seiberg:1994pq}, for a review see~\cite{Intriligator:1995au}. Interestingly there are various phases as the number of flavors $F$ is increased with respect to the number of colors $N$, and the IR dynamics is different in these various SUSY phases. It has been recently shown~\cite{Murayama:2021xfj,Csaki:2022cyg}, that SUSY QCD with AMSB has a vacuum with QCD-like chiral symmetry breaking $SU(F)\times SU(F)\to SU(F)_V$ (at least as a local vacuum) for all flavors where the original SUSY theory is asymptotically free ($F\leq 3N$)~\footnote{For other recent work applying AMSB to exact SUSY results see~\cite{Csaki:2021aqv,Csaki:2021jax,Csaki:2021xhi,Csaki:2021xuc,Murayama:2021rak,Bai:2021tgl,Kondo:2022lvu,Luzio:2022ccn,Ciambriello:2022wmh}. }. It is therefore possible to ask what the resulting $\eta'$ potential and chiral Lagrangian look like. The $\eta'$ potential, as well as the chiral Lagrangian, are (mostly) calculable within the scope of  AMSB QCD theories, except for some ${\cal O} (1)$ K\"ahler potential coeffecients for some specific number of flavors. 

Throughout our calculations we will assume $m_Q \ll m \ll \Lambda$, where $m$ is the SUSY breaking mass scale and $m_Q$ is the scale of the quark masses (that are added to the superpotential). This choice leaves squarks and gluinos below the strong coupling scale $\Lambda$ and makes them participate in the strong dynamics. Hence these theories do not truly have the same dynamics as ordinary QCD, but the massless spectrum is indeed just that of QCD, with the squarks and gluinos picking up an AMSB mass at one loop. Ordinary QCD would correspond to taking $m>\Lambda$, a limit we cannot take since there may be phase transitions occurring. Nevertheless we find some quite remarkable and unexpected results, which may serve as lessons for the dynamics of ordinary QCD as well: 

\begin{itemize}
\item The dynamics responsible for the $\eta'$ potential (and hence confinement and chiral symmetry breaking) is strongly dependent on the number of flavors $F$.

\item For $|F-N|>1$ there is indeed a branched structure for the $\theta$-dependent part of the potential, given by a function $V(\theta/|F-N|)$. The origin of the branches are gaugino condensation in the unbroken subgroup for $F<N$ and in the dual gauge group for $F> N+1$. 

\item For $F=N-1,N,N+1$ there is no branched structure for the potential, corresponding to the fact that the entire $\theta$-dependence is consistent with being generated by ordinary instantons.

\item While the $\eta'$ mass does vanish in the large $N$ limit as long as the number of flavors is fixed, if $F\propto N$ the $\eta'$ mass will not vanish, in agreement with the fact that the chiral anomaly also remains fixed for this case. 

\item {For $N\gg F$ the vacuum energy has the same $N$ dependence as in pure QCD without fermions, i.e. it is proportional to $N^2$. However, if both the number of flavors $F$ and the number of colors $N$ are large with $F\sim N \gg 1$ we find that the vacuum energy is proportional to $N^{3/2}$.}

\item {For generic $\theta$ the theory has a unique vacuum in the limit of equal quark masses. For $\theta = \pi$ the vacuum structure depends critically on $F$ and the quark masses $m_Q$. For $F=1$ there exists a critical value for the quark masses $m_{Q,0}\sim m /N$ below which there is a unique vacuum and CP remains unbroken. For $m_Q = m_{Q,0}$ the $\eta'$ is exactly massless and for $m_Q > m_{Q,0}$ there are two degenerate vacua and CP is spontaneously broken. 
For $F>1$ we always find doubly-degenerate vacua for all non-vanishing quark masses leading to spontaneous CP breaking and a first-order phase transition, in agreement with the findings of~\cite{Gaiotto:2017tne} for non-supersymmetric QCD.}
\end{itemize}

So what do we expect for ordinary QCD? In that case $F=N$, and it is not clear which (if any) of the large $N$ limits is the most relevant. Based on the lessons learned here one would expect that the light flavors can play an important role in the confining dynamics and are not negligible. However, the essence of the flavor dependence in the broken SUSY theories arises through the VEV of squarks, for example, if $N>F$, the group confining in the IR is reduced to $SU(N-F)$. Thus it is not clear whether a similarly strong $F$-dependence persists in the non-supersymmetric case.  On the other hand, the number of branches in SUSY QCD can also be understood as a simple consequence of the selection rules of its anomalous $U(1)_R$ symmetry (that plays a similar role as $U(1)_A$ in ordinary QCD). The number of flavors $F$ determines the size of the anomaly and through it the $\theta$-dependence of the potential, as we show in Section~\ref{sec:symmetry}. It is possible that this symmetry argument can be interpolated into the regime with large SUSY breaking relevant for ordinary QCD, though it is not immediately clear to us how to do it in a controlled way. The theory where this seems hardest to do is pure SUSY Yang-Mills, i.e. the $F=0$ limit of our models. SUSY Yang-Mills has $N$ branches, with the gluino condensate a function of $\theta/N$. After SUSY breaking via AMSB this translates into $N$  branches for the vacuum energy  as a function of $\theta$. This is a consequence of the $U(1)_R$ acting on the gluinos, and of its anomaly. In ordinary QCD at large $N$ we find exactly the same number of branches for the vacuum energy as a function of $\theta$~\cite{Witten:1978bc, DiVecchia:1980yfw}, even though there are no fermions present, and hence there is no trace of this symmetry or a possible analogous one.

Results qualitatively similar to ours have been obtained  by Dine, Draper, Stevenson-Haskins and Xu~\cite{Dine:2016sgq} by perturbing the exact SUSY results with soft squark and gluino masses (see also~\cite{Davies:2022ueb} for some comments). Here we use the specific form of AMSB which allows a complete mapping of the SUSY breaking terms in a UV insensitive way, which makes it possible to obtain the resulting confining potentials for the $F\geq N$ cases. 

The paper is organized as follows. In the first half of the paper we review the standard lore of the $\eta'$ in ordinary QCD. In Sections~\ref{sec:chiralLagrangian} and~\ref{sec:largeN} we outline how the $\eta'$ is introduced into the chiral Lagrangian and discuss possible origins of its non-perturbative potential: instantons or confinement dynamics. Then, using arguments from large $N$ QCD, we construct an improved potential for the $\eta'$ and investigate how it affects the axion mass in Section~\ref{sec:axionMass}. In the second half of the paper we move on to SUSY QCD with AMSB to study the origin of the $\eta'$ potential in a fully calculable framework. We summarize our findings from SUSY QCD in Section~\ref{sec:SUSYLessons}. After a short review of AMSB in Section~\ref{sec:AMSBReview} we use a spurion approach to identify the $\theta$ dependence of the scalar potential in the chiral Lagrangian in Section~\ref{sec:symmetry} before we systematically obtain and study the chiral Lagrangian for $F<N$ in Section~\ref{sec:FlessN}, $F=N,N+1$ in Section~\ref{sec:FeqN} and $F>N+1$ in Section~\ref{sec:FlargerN}. We conclude in Section~\ref{sec:conclusion}.

\section{General Structure of the Chiral Lagrangian\label{sec:chiralLagrangian}}
We are interested in the low energy dynamics of a confining $SU(N)$ gauge theory with $F$ quarks in its fundamental representation, and  always take $F$ below the conformal window. In analogy with QCD we assume that the quark masses are small compared to the confinement scale. 
In this Section, to set the notation, we introduce the well-known chiral Lagrangian that describes the lightest degrees of freedom of this theory, with an eye to the possible role of instantons and the computation of the $\eta^\prime$ and axion potentials.

At the classical level, when all quark masses vanish, there is a $U(F)_L\times U(F)_R=SU(F)_L\times SU(F)_R \times U(1)_L\times U(1)_R$ global symmetry,\footnote{Note that this isomorphism holds only locally. Globally $U(F) = (SU(F)\times U(1))/\mathbb{Z}_F$.} but quantum mechanically $U(1)_A$ is explicitly broken. The vacuum structure of the gauge theory is such that 
only the diagonal subgroup $U(F)_V$ is linearly realized, resulting in $F^2-1$ massless Golstone bosons, which we will simply call `pions' ($\pi^a$). The inclusion of identical quark masses explicitly breaks the $U(F)_L \times U(F)_R$ symmetry down to $U(F)_V$. The differences between quark masses can further break $U(F)_V$ to $U(1)^F$. Quark masses $m_q$  much smaller than the confinement scale $\Lambda$ can be considered a perturbation. In this limit the pions remain the lightest states of the theory, with masses suppressed by $m_q$. 

The vector $U(1)_V=U(1)_L+U(1)_R$ factor is identified with unbroken baryon number $U(1)_B$. 
The axial $U(1)_A=U(1)_L-U(1)_R$ is  anomalous, with the anomaly given by 
\begin{equation}\label{eq:anomaly}
\partial_\mu j^\mu_A = F \frac{g^2}{32\pi^2}  {\rm Tr}\, G \tilde{G} \,.
\end{equation}
Here we assumed that the anomaly is only due to the fundamental fermions of the $SU(N)$ gauge group, $g$ is the $SU(N)$ gauge coupling and ${\rm Tr}\, G \widetilde{G}\equiv \epsilon^{\mu\nu\rho\sigma}\sum_{a=1}^{N^2-1}G_{\mu\nu}^a G_{\rho\sigma}^a$ for a normalization of the $SU(N)$ generators of Tr$[T^a T^b]=2\delta^{ab}$. In the absence of the anomaly the Goldstone boson $\eta'$, associated to the $U(1)_A$ current, is massless. However, in the presence of the anomaly, the $\eta'$ is expected to be just another massive particle, much heavier than the pseudo-Goldstone pions. In particular, `t Hooft argued that instanton effects can explain the absence of a light $\eta'$ (solving the so called ``$U(1)$-problem")~\cite{tHooft:1976snw,tHooft:1986ooh}.

The general approach to capture the physics of the $U(1)_A$ breaking in the Chiral Lagrangian is to promote the $\theta$ parameter of the gauge theory, defined as 
\be \label{eq:thetaTerm}
\mathcal{L}\supset \theta\frac{g^2}{64\pi^2}{\rm Tr} G\widetilde{G}\, ,
\ee
to a spurion. Under a chiral rotation of the quarks
\be
\psi_j \to e^{i \varphi} \psi_j, \quad \psi_j^c \to e^{i \varphi} \psi_j^c\, , \quad j=1, ..., F\, ,
\ee
the path integral measure changes non-trivially~\cite{Fujikawa:1979ay,Fujikawa:1980eg}. This can be compensated by a shift of the $\theta$ angle:
\begin{equation}
\theta \to \theta + 2\, F \varphi \ . \label{thetashift}
\end{equation}
Assigning this transformation behavior to $\theta$ promotes it to a spurion and formally restores the $U(1)_A$ symmetry. Thus it can be used as a building block in the chiral Lagrangian to construct $U(1)_A$ invariant terms. The same can be done for the explicit breaking from the quark mass matrix $m_Q$, promoting it to a spurion of $SU(F)_L\times SU(F)_R\times U(1)_A$ in the usual way,
\be
m_Q \to e^{-2i \varphi} U_R m_Q U_L^\dagger\, .
\ee
where $U_L$ and $U_R$ are $SU(F)_L$ and $SU(F)_R$ transformations, respectively. 
To build the chiral Lagrangian we need to introduce the Goldstone fields $U$, which under the $SU(F)_L\times SU(F)_R\times U(1)_A$ global symmetries transform as 
\begin{equation}
U \to e^{2 i \varphi} U_L U U_R^\dagger
\end{equation}
and can be parametrized as
\begin{equation}
U = e^{i \eta'} e^{i \pi^a T^a}.
\end{equation} 
This is in accordance with the expectation that the $\eta'$ shifts under the axial symmetry as 
\begin{equation}
\eta' \to \eta'+ 2\varphi  \label{etapshift}\,.
\end{equation}
We have absorbed the decay constants $f_{\eta'}$ and $f_\pi$ into ${\eta'}$ and $\pi$, respectively, so that the meson fields are dimensionless. Note that $U\in U(F)_A = (SU(F)_A \times U(1)_A)/\mathbb{Z}_F$ which implies  the identification $(e^{i \pi^a T^a}, \eta') \sim (e^{-\tfrac{2\pi i}{F}} e^{i \pi^a T^a}, \eta'+\tfrac{2\pi}{F})$, i.e. the physical range of $\eta'$ in this normalization is $\eta'\in [0, \tfrac{2\pi}{F}]$ (see e.g.~\cite{Gaiotto:2017tne}).

The usual leading terms in the chiral Lagrangian can be written as 
\begin{equation}\label{eq:chiralLag}
{\cal L} = \frac{f_\pi^2}{4} {\rm Tr} \left[ (\partial_\mu U)^\dagger \partial^\mu U  \right] + \alpha \Lambda f_\pi^2\left( {\rm Tr} \left[m_Q U\right] + {\rm h.c.}\right)\,,
\end{equation}
where $\Lambda \simeq 4 \pi f_\pi$ is the dynamical scale of the gauge group, $\alpha$ is an $\mathcal{O}(1)$ number, $m_Q$ is the quark mass matrix, and we assumed that $f_{\eta'}=f_\pi$. Note that in the rest of the paper we show potentials for the mesons ($\eta^\prime$ and $\pi^a$) where they are not canonically normalized. 
We normalize the kinetic term only when showing the physical $\eta^\prime$ mass. This choice simplifies the spurion analysis of the $U(1)_A$ symmetry.

To reproduce the symmetry properties of the high-energy theory in our effective theory without fermions, we add to the above Lagrangian a term that breaks the $U(1)_A$ consistent with the spurion analysis above. The simplest possibility appears to be 
\begin{equation}
{\cal L}_{\rm inst} = b \Lambda^2 f_\pi^2 e^{-i\theta} \det U +{\rm h.c.}\,, \label{wrongterm}
\end{equation}
where $b$ is an unknown dimensionless coefficient. This term  breaks the axial symmetry explicitly, which is however restored if we promote $\theta$ to a spurion. Eq.~\eqref{wrongterm} may correspond to an ordinary instanton because it is proportional to $e^{-i\theta}$, the hallmark of 1-instanton effects, which we will expand on in the next Section. The resulting potential for the $\eta'$ is 
\begin{equation}\label{eq:QCDEtap}
V_{\eta'} =  -2 b \Lambda^2 f_\pi^2 \cos (\theta - F \eta') \, ,
\end{equation}
a function which is explicitly $2\pi$ periodic in $\theta$ without branch cuts or singularities. In the absence of quark masses, given the transformations Eq.~\eqref{thetashift} and Eq.~\eqref{etapshift} any potential term can only depend on the $U(1)_A$ invariant combination $\theta- F\eta'$.

To analyze the vacuum structure of the theory (and the axion mass) one can integrate out the $\eta'$ and after that the pions. Since the $\eta'$ is much heavier than the pions, at leading order its vev  is determined from Eq.~\eqref{eq:QCDEtap}, $\eta' = (\theta + 2 k \pi )/F$, where $k$ is an arbitrary integer. In the following we will restrict ourselves to the physical field range $\eta' \in [0,\tfrac{2\pi}{F}]$ and therefore set $k=0$. We can also assume that the quark mass matrix has only one overall phase $\theta_q$, i.e. $m_Q= e^{i\theta_q}m_q$ (which can always be achieved by a suitable $SU(F)_L\times SU(F)_R$ rotation). Hence the potential for the lightest pseudo-Goldstone bosons can be obtained from Eq.~\eqref{eq:chiralLag} and is given by 
\begin{equation}\label{eq:QCDGBMassTerm}
V_\pi=-\alpha \Lambda f_\pi^2 e^{i \bar{\theta}/F} {\rm Tr} (m_q e^{i \pi^a T^a}) +{\rm h.c.}\,,
\end{equation}
where $\bar\theta = \theta + F \theta_q$ 
is the usual physical combination that can be measured. To find the $\bar{\theta}$ dependence one needs to minimize the potential with respect to the neutral Goldstone bosons. For $F$ flavors there will be $F-1$ neutral Goldstones corresponding to the Cartan sub-algebra of $SU(F)_A$. The Cartan sub-algebra is generated by the $F-1$ generators, $t^1, \ldots , t^{F-1}$, that can be simultaneously diagonalized. The resulting potential is 
\begin{equation}\label{eq:QCDGBPotential}
V_{\pi^0}= -2 \alpha \Lambda f_\pi^2 \sum_{i=1}^{F} m_i \cos \left(\frac{\bar \theta}{F} +\sum_{j=1}^{F-1} t^j_i \pi^j \right)
\end{equation}
where $m_i$ is the i$^{th}$ diagonal element of the quark mass matrix and $t^j_i$ is the i$^{th}$ diagonal element of the j$^{th}$ Cartan generator. Note that despite its appearance Eq.~\eqref{eq:QCDGBPotential} is $2\pi$ periodic in $\bar{\theta}$. A shift $\bar{\theta}\rightarrow \bar{\theta} + 2\pi$ can be compensated by a redefinition of the GB fields $\pi^j$ and the $2\pi$ periodicity of the cosine. Clearly if any $m_i=0$ one can simply set the remaining $F-1$ arguments of the cosines to zero and reabsorb $\bar\theta$ into the VEVs of the neutral mesons. However if all $m_i$'s are non-zero one needs to minimize the potential of the sum of cosines and the value at the minimum will be $\bar\theta$-dependent, leading to a non-vanishing axion mass.

For example, for $F=2$ the potential is 
\begin{equation}\label{eq:2FlavorPotGB}
V_2= -2 \alpha \Lambda f_\pi^2 \left[m_u \cos \left(\frac{\bar\theta}{2} + \pi^0\right) +m_d \cos \left(\frac{\bar\theta}{2} - \pi^0\right) \right]\,.
\end{equation}
The minimum of $V_2$ is given by 
\begin{equation}\label{eq:2FlavorPot}
V_{\rm min}=-2 |\alpha| \Lambda f_\pi^2 \sqrt{m_u^2+m_d^2 +2 m_u m_d \cos \bar\theta}\,.
\end{equation}
For $F=3$ the potential is
\begin{equation}\label{eq:3FlavorPot}
\begin{split}
V= -2 \alpha \Lambda f_\pi^2 \left[m_u \cos \left(\frac{\bar\theta}{3} + \pi^0+ \frac{\eta}{\sqrt{3}}\right) +m_d \cos \left(\frac{\bar\theta}{3} - \pi^0+\frac{\eta}{\sqrt{3}}\right) \right.\\
\left. +m_s \cos \left(\frac{\bar\theta}{3} - \frac{2 \eta}{\sqrt{3}}\right)\right]\,.
\end{split}
\end{equation}
The equations for $\pi^0 ,\eta$ have to be minimized numerically. 
Note that the potentials in Eq.~\eqref{eq:2FlavorPotGB} and Eq.~\eqref{eq:3FlavorPot} can be made manifestly $2\pi$ periodic in $\bar{\theta}$ by shifting $\pi^0$ and $\eta$. In the $F=2$ case this shift takes the form $\pi^0\rightarrow \pi^0 -\frac{\bar{\theta}}{2}$.
 
\section{Instanton vs. condensates: large $N$ limit and branched potential\label{sec:largeN}}

The chiral Lagrangian term Eq.~\eqref{wrongterm}, has the characteristic form of a one-instanton effect. 
The action of a single instanton is $S_I=8\pi^2/g^2$, and an instanton always appears  with an $e^{\pm i\theta}$ factor, because it has winding number one for an instanton and minus one for an anti-instanton. This means that a one-instanton effect is always proportional to 
\begin{equation}
  e^{-8\pi^2/g^2\pm i\theta}  \ .
\end{equation}
In supersymmetric theories (as we will see in the second half of this paper) it is customary to introduce a complex (``holomorphic") coupling constant $\tau = \frac{4\pi i}{g^2} + \frac{\theta}{2\pi}$
where $g$ is the gauge coupling. The instanton effect is then proportional to $e^{2\pi i \tau}$. One important takeaway is that instanton effects will always involve an explicit $e^{\pm i n \theta}$ factor, where $n$ is an integer, giving rise to the explicit breaking of the axial symmetry. 

Another important quantity to consider is the dynamical scale of the theory, the generalization of $\Lambda_{\rm QCD}$. To one loop order it is defined as 
\begin{equation}
    \Lambda = \mu e^{-\frac{8\pi^2}{b_0 g^2(\mu )}},
\end{equation}
where $\mu$ is an arbitrary scale, and $b_0$ is the one loop beta function coefficient. One can easily show that this scale is RGE invariant to one loop order. This shows, that instanton effects are proportional to $\Lambda^{b_0}$, which is usually called the instanton factor. It is now clear, that it may be useful to define a holomorphic version of this dynamical scale that also incorporates the $\theta$ dependece of the one-instanton effect, which is simply
$\Lambda^{b_0}= \mu^{b_0} e^{2\pi i\tau}$. One of the great advantages of this holomorphic scale is that it carries a spurious charge under the anomalous axial symmetry, and it can be used as the spurion for the breaking of the axial symmetry via anomalies. For a more detailed discussion of the definition of the dynamical scale (especially in supersymmetric theories) see Appendix~\ref{app:holomScale}.

Witten and Veneziano  \cite{Witten:1978bc,Veneziano:1979ec,Witten:1980sp} pointed out that the situation regarding the $\eta'$ potential might not be as simple as adding the one-instanton motivated effective operator in Eq.~\eqref{wrongterm}. The best way to see the possible issue is by considering the large $N$ limit of the theory, keeping the `t Hooft coupling $\lambda=g^2N$ fixed. The chiral anomaly (assuming the number of flavors is held fixed) vanishes in this limit
\begin{equation}
\partial_\mu j^\mu_A \sim  F \frac{g^2}{8\pi^2}  {\rm Tr} G \tilde{G} \sim \frac{\lambda}{8\pi^2} \frac{F}{N}   {\rm Tr} G \tilde{G} \to 0 
\label{eq:anomalyF}
\end{equation}
and $U(1)_A$ is restored. Hence the expectation is that $\eta'$ can be treated on the same footing as all other mesonic GBs in this limit, i.e. its mass vanishes for massless quarks. However the type of instanton-inspired term, Eq.~\eqref{wrongterm}, that we have used in the previous Section does not go to zero for $N\to \infty$, since $\Lambda$ is fixed. In the large $N$ limit it is unlikely to capture the correct physics responsible for the $\eta'$ mass. A naive argument would suggest that all instanton effects should vanish in the large $N$ limit, since the instanton action $e^{-8\pi^2/g^2}\propto e^{-N}$, however this may not be correct due to infrared divergences and the growth of the number of zero modes one needs to integrate over. We will in fact see later cases when there are finite instanton effects even at large $N$, unsuppressed by $e^{-N}$.

Another convincing argument by Witten that Eq.~\eqref{wrongterm} is not the leading contribution to the potential, comes from
considering the vacuum energy of the theory. In pure QCD (without fermions), the vacuum energy is proportional to $N^2$---scaling with the number of gluons in the theory---and has a non-trivial dependence on $\theta$ of the form~\cite{Witten:1979vv}
\begin{equation}\label{eq:vacuumTheta}
E(\theta ) = N^2 f(\theta / N )\, ,
\end{equation}
for some function $f$. This is motivated by exact results in two dimensional models and by the fact that it reproduces the expectation that the $\eta'$ mass vanishes in the $N\rightarrow \infty$ limit, as we will verify momentarily. The vacuum energy, like all physical quantities, should be $2\pi$-periodic in $\theta$. In order to achieve this despite the dependence on $\theta$ only through $\theta /N$ Witten proposed that the potential is in fact continuous but not-smooth in $\theta$ with multiple branches. We will return to the form of this potential shortly.

Assuming that the small quark masses do not change the underlying dynamics and have only a small effect on the potential, then the potential for $\eta'$ can be deduced from the pure QCD vacuum energy $E(\theta)$. Using the $U(1)_A$ symmetry, where $\theta$ is promoted to a spurion transforming as in Eq.~\eqref{thetashift} under the $U(1)_A$, the potential with vanishing quark masses would be of the form
\begin{equation}
    V_F (\theta , \eta') = V_{\rm pure~QCD} (\theta - F \eta').
\end{equation}
Using the expression in Eq.~\eqref{eq:vacuumTheta} for the vacuum energy, this leads directly to the Veneziano-Witten formula~\cite{Witten:1978bc,Veneziano:1979ec} for the $\eta'$ mass\footnote{Note that $\frac{d^2}{d\eta'^2} V_{\rm pure~QCD} (\theta - F \eta') = \frac{F^2}{f_\pi^2} \frac{d^2}{d\theta^2} V_{\rm pure~QCD} (\theta - F \eta')$. The actual prefactor is achieved after the $\eta'$ kinetic term is canonically normalized.} 
\begin{equation}
m_{\eta'}^2 = \frac{2F}{f_\pi^2} \left.\frac{d^2 E}{d\theta^2}\right|^{{\rm pure\ QCD}}_{\theta=0}\, .
\end{equation}
From Eq.~\eqref{eq:vacuumTheta} it is apparent that $d^2/d\theta^2 \left.E(\theta)\right|_{\theta = 0} \sim N^0$ which together with the $N$ scaling of the pion decay constant $f_\pi \sim \sqrt{N}$ implies that $m_{\eta'}^2 \sim 1/N$, as expected in the large $N$ limit. This further justifies the ansatz for the vacuum energy in Eq.~\eqref{eq:vacuumTheta}.

In order to incorporate these results in the chiral Lagrangian we have to modify the term for the $\eta'$ mass. Requiring that the potential has $N$ branches as Witten argued, we suggest that instead of Eq.~\eqref{wrongterm} the proper term should rather be of the form
\begin{equation}
{\cal L}_{\eta'} = N \Lambda^2 f_\pi^2 (e^{-i\theta} {\rm det}\, U)^{1/N} +h.c.
\label{eq:branches}
\end{equation}
This potential correctly reproduces the expected scaling $m_{\eta'}^2 \sim 1/N$, and matches what we will find in the SUSY case below.\footnote{Note that one may instead use a term $1/N (-i\log\det U -\theta )^2$ which is essentially just a pure $\eta'$ mass term $1/N (F \eta' - \theta )^2$. Expanding Eq.~(\ref{eq:branches}) gives exactly this mass term to leading order, while the quartic $\eta'^4$ is suppressed by $N^4$ as expected.} 

The form of this potential has several important consequences.
First, the dynamics of the $\eta'$ mass does not actually directly originate from an instanton effect. Instanton terms should always be proportional to $e^{i n \theta}$ with $n$ integer. 
Second, the non-analytic form of Eq.~(\ref{eq:branches}) implies that the vacuum structure of pure QCD is, as already anticipated, non-trivial  with various branches. This ensures that physics remains $2\pi$ periodic in shifts of $\theta$. For example, the pure QCD potential could be of the form
\begin{equation}\label{eq:pureQCDPotential}
V(\theta ) = \min_k\left[\, -\frac{2N^2}{(4\pi)^2} \Lambda^4 \cos \left(\frac{\theta + 2\pi k}{N}\right)\right] , \ \  k =0,\ldots , N-1
\end{equation}
and would satisfy the conditions on the $\eta'$ mass and the periodicity of the theory in $\theta$. In this case one has $N$ different branches. 
\begin{figure}[t!]
    \centering
    \subfigure{\includegraphics[width=0.48\textwidth]{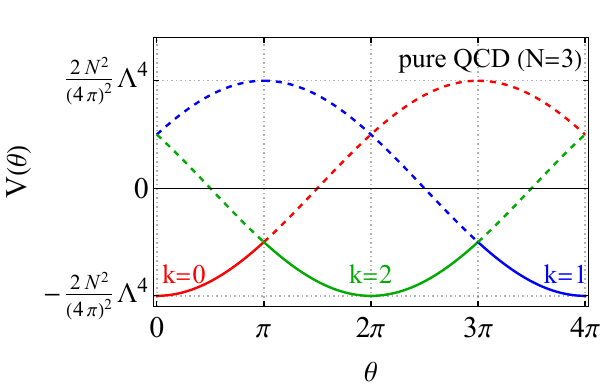}}
    \hfill
    \subfigure{\includegraphics[width=0.47\textwidth]{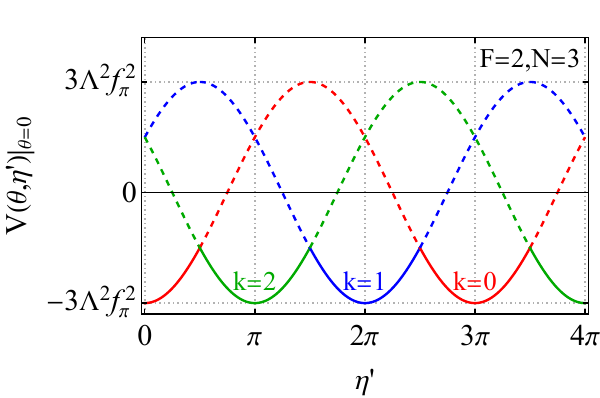}}
    \caption{Potential for $N=3$ pure QCD as given in Eq.~\eqref{eq:pureQCDPotential} (left) and $\eta'$ according to Eq.~\eqref{eq:QCDactual} for $N=3, F=2$ (right) along $\theta=0$. The three branches are depicted in different colors. The actual potential is the lower envelope of the branches, i.e. the solid curve.}
    \label{fig:etapPotential}
\end{figure}
It is important to point out that this branched structure for the potential is a well-informed guess in ordinary QCD. The arguments reviewed above and in~\cite{Witten:1980sp} lead to postulating a dependence of the vacuum energy of the form $V=V(\theta/N)$. 
In the following we see how these branches emerge explicitly in SUSY QCD.

Once fermions are introduced, the $\theta$-dependence will change to $\theta \to \theta -\eta' F$, and the potential in the chiral Lagrangian responsible for the $\eta'$ mass will be of the form 
\begin{equation}
V(\theta , \eta') =  \min_k\left[\, -2N \Lambda^2 f_\pi^2 \cos \left(\frac{\theta - F\eta' +  2\pi k}{N}\right)\right] , \ \  k =0,\ldots , N-1\, .
\label{eq:QCDactual}
\end{equation}
The potentials in Eq.~\eqref{eq:pureQCDPotential} and Eq.~\eqref{eq:QCDactual} are plotted in Fig.~\ref{fig:etapPotential} for $N=3$. The solid curve gives the full potential, while the dashed lines show the contribution to the potential of the different branches. The potential is not smooth, but remains periodic. The true minimum of the energy is for $\theta=0$. 

The $\eta'$ will adjust to the minimum of the potential so as to cancel the $\theta$-dependence, in essence itself acting like a heavy QCD-scale axion. 
This will wash out the presence of the various branches of pure QCD, with the only remnant being the value of the $\eta'$ VEV
\begin{equation}
\langle \eta' \rangle = \frac{\theta +2\pi k}{F} \,.
\end{equation}
It is important to note that $\eta'$ is an angular variable with periodicity $\tfrac{2\pi}{F}$ in our normalization. This implies that there is a single minimum in the physical field range $\eta'\in [0,\tfrac{2\pi}{F}]$, which corresponds to the choice $k=0$.
In the absence of quark masses (explicit breaking terms) the $\theta$ dependence completely disappears, as expected. Once quark masses are added, the $\theta$-dependence resurfaces through the $\theta$-dependence of the $\eta'$ VEV (which now is just an overall phase of the $U$ matrix); see Eq.~\eqref{eq:QCDGBPotential}. However the story is still not finished: the light pseudo-Goldstone bosons themselves act as axions and would like to cancel the remaining $\theta$-dependence of the Lagrangian. For $F$ quark masses there are only $F-1$ neutral Goldstone bosons, and one cannot fully cancel all the $\theta$-dependence of the Lagrangian, hence the need for the usual axion that can cancel the remaining $\theta$-dependence. If at least one of the quark masses vanishes then there are enough neutral Goldstone bosons to completely cancel the $\theta$ dependence, hence the $m_u=0$ solution of the strong CP problem. 

We have seen that the most likely dynamical origin for the $\theta$-dependence of the QCD potential is not actually a direct instanton effect, but rather the confining dynamics that gives rise to the various condensates of QCD. A nice heuristic picture of Di Vecchia and Veneziano~\cite{DiVecchia:1980yfw} starts with the fact that the low-energy theory should contain an $\eta' {\rm Tr} G \tilde{G}$ term to reproduce the chiral anomaly of theories with fermions. Once confinement happens this term can be thought of as a mixing between the $\eta'$ and a pseudo-scalar glueball whose interpolating field is ${\rm Tr} G\tilde{G}$. The glueball should also have a direct mass term generated by confinement. In this picture the mixing between the $\eta'$ and the pseudo-scalar glueball is the origin of the $\eta'$ mass. In the supersymmetric theory we will often find gluino condensation as the origin of confinement, the $\eta'$ mass and the various branches of the theory. 
 
\section{The axion mass}\label{sec:axionMass}

Let us now investigate how our previous discussion affects the potential of the axion. To do this we have to assume that there is a second chiral $U(1)$ Peccei-Quinn (PQ) symmetry, which is non-linearly realized at some high scale $f_a$, and is  anomalous under QCD. The PQ symmetry acts as a shift symmetry on the resulting Goldstone boson, the axion: $a \to a + \varphi$. The anomaly explicitly breaks this symmetry, and to formally restore it we can promote the $\theta$-angle to a spurion
\begin{equation}
    a\to a + \varphi , \qquad \theta \to \theta - n \varphi\,,
\end{equation}
where $n$ is the anomaly coefficient of $U(1)_{PQ}$ under QCD. This implies that the QCD potential now depends on the combination $\theta + n a - F\eta'$. The QCD potential gives a mass to one combination of $\eta'$ and $a$, and the orthogonal combination remains massless. Since the dimensionless $\eta'$ is actually suppressed by $f_\pi$ while the dimensionless $a$ by $f_a$, and $f_a \gg f_\pi$, the massive field is mostly composed by the $\eta'$, and the massless one to a good approximation is the axion. 

For concreteness let us consider the potential on the $k^{\rm th}$ branch
\begin{equation}
    V_k(\eta' , a, \pi^j ) = -2N \Lambda^2 f_\pi^2 \cos\left( \frac{\theta -F \eta' + n a+2\pi k}{N}\right) -2\alpha \Lambda^2 f_\pi^2 \sum_{i=1}^F \frac{m_i}{\Lambda} \cos \left(\eta' +\theta_q +\sum_{j=1}^{F-1} t^j_i  \pi^j\right)
    \label{eq:etappotential}
\end{equation}
Integrating out the $\eta'$ gives to leading order in $m_i/\Lambda$\footnote{Implicitly we assume $\alpha m_i \ll \tfrac{F}{N} \Lambda$.}
\begin{equation}
 \eta' = \frac{1}{F} (\theta + n a + 2\pi k) \ .
\end{equation}
As expected the $\eta'$ adjusts to cancel the QCD potential, and to leading order washes out all the effects of the various branches. The axion potential, to leading order is then 
\begin{equation}
    V_{a}= -2\alpha \Lambda^2 f_\pi^2 \sum_{i=1}^{F} \frac{m_i}{\Lambda}  \cos\left( \frac{\bar{\theta} + a n}{F} +\sum_{j=1}^{F-1} t_i^j \pi^j\right)
\label{eq:axionpot}
\end{equation}
where $\bar{\theta}= \theta+ F\theta_q$ is the physical $\bar{\theta}$ and we used $\eta'=\eta'+2\pi k/F$.\footnote{One should not try to minimize the $\eta'$ potential in (\ref{eq:etappotential}) to higher order in $m_i/\Lambda$. Including the shift in the $\eta'$ VEV due to the quark masses will have an effect on the axion potential equivalent to considering a term suppressed by higher powers of $m_i/\Lambda$ in the chiral Lagrangian of the form $- \alpha^2 f_\pi^2 \frac{N}{8 F^2} \left[ \text{Tr}\left( m_Q U - U^\dagger m_Q^\dagger \right) \right]^2\,$.}

This discussion also clarifies that while the axial anomaly is related to the generation of the QCD contribution to the axion mass, it is {\it not} IR instantons that directly contribute to the axion potential. Thus attempts at trying to draw instanton diagrams representing 't Hooft operators in order to explain the usual axion mass formula are futile. This of course does not mean that there could not be {\it additional} contributions from small instantons much above the QCD scale. There are indeed many models for that, using various modifications of the QCD dyanmics in the UV to obtain such terms (see e.g.~\cite{Holdom:1982ex,Holdom:1985vx,Flynn:1987rs,Agrawal:2017ksf,Csaki:2019vte,Gherghetta:2020keg}). 

Finally let us discuss a simple method to obtain a closed form expression of the axion mass for an arbitrary number flavors, to leading order in $m_i/\Lambda$ and $f_\pi/f_a$. 
With the axion as a dynamical field in (\ref{eq:axionpot}), it is trivial to find the minimum of this potential: the axion will just cancel $\bar\theta$ while all the pions will have a vanishing VEV. Hence finding the mass matrix is very simple, it is just a sum of pure quadratic terms 
\begin{equation}
    V_{a}= \alpha \Lambda^2 f_\pi^2 \sum_{i=1}^F \frac{m_i}{\Lambda} \left(\frac{a n}{F} +\sum_{j=1}^{F-1} t_i^j \pi^j\right)^2\,.
\end{equation}
Integrating out the pions (which are much heavier than the axion) we directly obtain the expression for the axion mass for an arbitrary number of flavors\footnote{For more details see Appendix~\ref{app:axionMass}.}
\begin{equation}
    m_a^2 =  \alpha \Lambda n^2 \frac{f_\pi^2}{f_a^2}\left(\sum_{i=1}^F m_i^{-1}\right)^{-1}\,.
\end{equation}
As expected, if any of the quark masses vanish, the axion mass will vanish too. 
The coefficient can be related to the pion masses by using the relation $\sum_i^{F-1} m_{\pi_i}^2 = 4 \alpha \Lambda \frac{F-1}{F}\sum_i^{F} m_i$ 
to arrive at the axion mass
\begin{equation}
m_a^2 = 
 \frac{n^2 F}{2(F-1)} \frac{f_\pi^2}{f_a^2}\frac{{\rm Tr}\, m_{\pi}^2 }{{\rm Tr}\,m_q {\rm Tr}\, m_q^{-1}}\,.
\end{equation}
 For $F=2$ we get the usual expressions 
\begin{equation}
    m_a^2 = 2\alpha \Lambda n^2 \frac{f_\pi^2}{f_a^2}
    \frac{m_u m_d}{m_u+m_d}=n^2 m_\pi^2 \frac{f_\pi^2}{f_a^2}
    \frac{m_u m_d}{(m_u+m_d)^2}\ .
\end{equation}

\section{Lessons for the Chiral Lagrangian from Supersymmetric QCD with AMSB: Summary of Results}\label{sec:SUSYLessons}
Now that we have reviewed the standard lore about the dynamics leading to the $\eta'$ and axion masses, we are ready to present our results for the analogous quantities in the supersymmetric extensions of QCD, where a small amount of supersymmetry breaking is introduced via anomaly mediation (AMSB)~\cite{Randall:1998ra,Giudice:1998xp,Luty:1999qc,Pomarol:1999ie}. We will start with the exact vacuum of SUSY QCD 
and then introduce SUSY breaking via AMSB. As explained in~\cite{Murayama:2021xfj} the effect of AMSB will generate a mass for the squarks and gluinos proportional to the amount of SUSY breaking denoted by $m$. To mimick ordinary QCD, we will also introduce quark masses $m_Q$ in the superpotential, and consider  the limit $m_Q\ll m\ll \Lambda$. This will allow us to find the chiral Lagrangian of this QCD-like theory, and in particular identify the potential of the $\eta'$, as well as the $\theta$-dependence of the resulting vacuum energy. Before detailing our calculations, we will give a brief summary of our major results:
\begin{itemize}
    \item We verify Witten's conjecture~\cite{Veneziano:1979ec,Witten:1978bc, Witten:1980sp} that the $\eta'$ potential has a branched structure. We also verify that for large $N$, but small $F$, the periodiciy of each branch is $2\pi N$.
    \item However, for general number of flavors, the confining potential changes qualitatively. In particular, the periodicity is given by $2\pi |N-F|$ and the potential is not simply the QCD potential with $\theta$ replaced by $\theta - F \eta'$. 
    \item As Witten postulated~\cite{Witten:1978bc, Witten:1980sp}, the origin of the branches lies in the dynamics responsible for confinement. In the case of our almost supersymmetric theory this dynamics is gaugino condensation in the unbroken gauge group (or for $F>N+1$ in the unbroken dual gauge group). 
    \item Within the models considered here the dynamical origin of the $\eta^\prime$ potential at large $F$ comes from the breaking of the gauge group from $SU(N)$ to $SU(N-F)$ via the $F$ squark VEVs  for $F<N$, while for $F>N$ the presence of flavors changes the IR dynamics to that of a dual $SU(F-N)$ gauge group which in turn will have its own confining dynamics via gaugino condensation.
    \item For $F=N-1,N,N+1$ we find no branch structure of the $\eta'$ potential. In this case the dynamics leading to the vacuum structure is indeed consistent with being an instanton effect. 
    \item The $\eta'$ mass does not vanish in the large $N$ limit if also $F \propto N$. This is most easily seen in the $F=N-1,N,N+1$ special cases where there is no branch structure to begin with, but also applies to the cases with gaugino condensation. 
    The non-vanishing of the $\eta'$ mass in the large $N$ limit is not too surprising, since the anomaly is also non-vanishing for $F\propto N$.  
    \item The vacuum energy scales as $N^2$ for $N\gg F$, a scaling which is consistent with the number of degrees of freedom (gluons and gluinos) in the theory. However, this scaling gets reduced to $N^{3/2}$ if also the number of flavors is large, i.e. $F\sim N \gg 1$.
    \item In the limit of equal quark masses the theory has a unique vacuum for generic $\theta$. For $\theta =\pi$ we find that for $F=1$ and small quark masses below a critical value $m_{Q,0}$ there is still a unique vacuum and CP remains unbroken. At the critical value, i.e. for $m_{Q}=m_{Q,0}$ the $\eta'$ is massless and for $m_Q > m_{Q,0}$ there are degenerate vacua and CP is spontaneously broken. For $F>1$ there are always doubly-degenerate vacua for all non-vanishing quark masses leading to spontaneous CP breaking and a first-order phase transition. This agrees with the findings of~\cite{Gaiotto:2017tne} for non-supersymmetric QCD.
\end{itemize}
A typical result that illustrates the above points is the form of the $\eta'$ potential for $F<N$ on the $k^{th}$ branch:
\begin{equation}
V_k =  -f(N,F) \left(\frac{m}{|\Lambda |}\right)^{F/N} m |\Lambda |^3 \cos\left( \frac{F \eta' +\theta +2 \pi k}{N-F}  \right)+ {\cal O} \left(\frac{m_Q}{m}\right)
\end{equation}
where $f(N,F)$ is a numerical function of $N,F$. We can see nicely that the branch structure is determined by $N-F$, in particular for $N-F=1$ there will be no branches, and the dynamical origin of this potential is a pure instanton term.  For $F<N-1$ this potential originates from gaugino condensation in an $SU(N-F)$ subgroup, which leads to $N-F$ branches. We can also see that for $F\propto N$ the $\eta'$ mass does not go to zero\footnote{At least as long as $f(N,F)$ does not vanish in that limit, which we will turn out to be the case in the explicit calculation}. 

Once the $\eta'$ is integrated out, we will obtain potentials analogous to Eq.~(\ref{eq:QCDGBPotential}). The typical form will be 
\begin{equation}\label{eq:AMSBPotential}
V =  -g(N,F) \left(\frac{m}{|\Lambda |}\right)^{F/N} |\Lambda |^3 \sum_{i=1}^{F} m_i \cos\left( \frac{\bar\theta}{F} + \sum_{j=1}^{F-1} t^j_i \pi^j \right)\,,
\end{equation}
where the branch structure due to the strong dynamics is washed out.
This is exactly like in ordinary QCD, and the resulting axion mass (if a physical axion is introduced) will have a structure similar to that in ordinary QCD.

\section{Review of AMSB}\label{sec:AMSBReview}

First we will quickly review the most important formulae needed for our calculations, for more details see~\cite{Murayama:2021xfj,Csaki:2022cyg}. We will assume throughout that an ${\cal N}=1$ supersymmetric extension of QCD (SUSY QCD) obtains soft breaking terms via anomaly mediation~\cite{Randall:1998uk,Giudice:1998xp,Arkani-Hamed:1998mzz,Luty:1999qc,Pomarol:1999ie}. For other appraoches to perturbing exact results with soft breaking terms see~\cite{Evans:1995ia,Alvarez-Gaume:1996vlf,Konishi:1996iz,Cheng:1998xg,Arkani-Hamed:1998dti,Abel:2011wv,Cordova:2018acb}. The mass scale of supersymmetry breaking is denoted by $m$, and AMSB will give rise to SUSY breaking effects wherever there is a source of violation of scale invariance. This can be tracked by including a conformal compensator $\Phi$ which will obtain an $F$-term set by the SUSY breaking scale $m$, $\Phi = 1 +\vartheta^2 m$. This will result in two types of SUSY breaking terms. There will be a tree-level potential term for the squarks of the form~\cite{Csaki:2022cyg} 
\begin{equation} \label{eq:VtreeAMSB}
V_{\text {tree }}=\partial_i W g^{i j^*} \partial_j^* W^*+m^* m\left(\partial_i K g^{i j^*} \partial_j^* K-K\right) +m\left(\partial_i W g^{i j^*} \partial_j^* K-3 W\right)+h . c .\,,
\end{equation}
generated whenever the K\"ahler potential is not quadratic or the superpotential not cubic. The potential is assumed to be along the D-flat direction.  Here
$ g_{i j^*} = \partial_i \partial_{j}^\ast K$ is the K\"ahler metric and $g^{i j^*}$ its inverse. These terms play a crucial role in finding the correct vacuum structure.

Due to the RGE running AMSB also generates gaugino and squark masses at the loop-level~\cite{Murayama:2021xfj}
\begin{equation} \label{susybreakingmasses}
m_\lambda = \frac{g^2}{16\pi^2} (3N - F)m\,,\quad m_{\widetilde{Q}}^2 = \frac{g^4}{(8\pi^2)^2} 2 C_i (3N -F) m^2\,,
\end{equation}
with $C_i = (N^2-1)/(2N)$. This sets up the UV boundary condition where all squark mass squares are positive as long as the gauge group is asymptoticall free, i.e. $F<3N$. Hence with AMSB we are investigating a QCD-like theory where the massless spectrum exactly matches that of QCD, however we also have the superpartners present at the massive level, and still well below the scale of strong dynamics. These squarks and glunios will still participate in the strong dynamics, hence the theory is not exactly QCD. One can connect this theory to QCD by taking the $\Lambda \ll m \to \infty$ limit, however it is not clear whether a phase transition occurs when $m$ goes above $\Lambda$. We will not be trying to take the $m>\Lambda$ limit, but rather investigate the dynamics of the QCD-like theory with the additional superpartners. Note that in the limit $N\gg F$ the physical masses $m_\lambda, m_{\widetilde{Q}}, m_{\widetilde{\bar{Q}}} \propto (g^2 N) m$ depend only on the constant combination $g^2 N =$ const., hence the physical masses are finite in the large $N$ limit if $m$ has no dependence on $N$.

\section{A Spurion Argument for the $\theta$-dependence}
\label{sec:symmetry}

Our main results on the $\theta$-dependence of the vacuum energy derived in Sections~\ref{sec:FlessN}, \ref{sec:FeqN} and \ref{sec:FlargerN},
can be understood as simple consequences of the anomalous $U(1)_R$ symmetry of SUSY QCD. While these arguments are well-known and are already discussed as early as in~\cite{Affleck:1983mk} we find it instructive to show explicitly how they can be used to find the $\theta$-dependence of the potential. 

 Supersymmetric theories often contain a peculiar type of chiral symmetry, called the R-symmetry, under which the various elements of a supermultiplet have different charges. The best way to explain this symmetry is to assign R-charge +1 to the $\vartheta$ coordinate of superspace (which is not to be confused with the $\theta$ angle). Since a chiral superfield has an expansion of the form $\Phi = \varphi (y) +\vartheta \psi (y) +\vartheta^2 F(y)$, this implies that if the scalar has R-charge $r$ then the fermion has R-charge $r-1$, and the F-component has R-charge $r-2$. Since the contribution of a superpotential to the Lagrangian is of the form $\int d^2\vartheta W$, a superpotential term has to have R-charge 2, which usually determines the R-charges of the chiral superfields. Finally, the gauge kinetic term can also be written in the chiral form $\int d^2\vartheta W_\alpha W^\alpha$ where $W_\alpha = \lambda_\alpha + \ldots$ is the vector superfield whose lowest component is the gaugino $\lambda_\alpha$. Hence a gaugino always has to have R-charge +1.

Let us first consider the case of pure super Yang-Mills (SYM) $F=0$. The R-symmetry in this case only acts on the gluinos, and is an axial rotation $\lambda \to e^{i\alpha}\lambda$  
which is anomalous under the gauge group. We can formally restore the anomalous symmetry by promoting $\theta$ to a spurion,
\be
\theta \to \theta + 2N\alpha \, ,
\ee
where $2N$ is the anomaly coefficient. Since physics is invariant under $\theta\to \theta+2\pi$, a $Z_{2N}$ discrete subgroup of $U(1)_A$, given by $\alpha=2\pi k/(2N)$, survives as a symmetry at the quantum level. This anomalous R-symmetry can be used to find the effective superpotential of the low-energy theory, assuming that pure SYM is confining with a mass gap (ie. gluinos and/or gluons condense and form massive glueballs and gluinoballs). The $\theta$-dependence of the effective superpotential can be immediately fixed from spurion analysis, as a superpotential term has to have R-charge 2 (and dimension 3), hence 
\be
W_{\rm eff}= c\, \mu^3 e^{i \frac{\theta}{N}}\,  \label{eq:N}
\ee
where $\mu$ is some dimensionful parameter. Since we are considering a supersymmetric theory whose superpotential has to be holomorphic, the above superpotential can be rewritten in terms of the holomorphic gauge coupling $\tau = \frac{4\pi i}{g^2} +\frac{\theta}{2\pi}$ as $W \propto e^{6 \pi i \tau / b_0}$ with $b_0=3N$ for SYM. Introducing the holomorphic dynamical scale $\Lambda=\mu e^{\frac{2\pi i \tau(\mu)}{b_0}}$ we find the usual form of the effective superpotential for SYM, $W_{\rm eff}= c^\prime \Lambda^3$. 
In order for physics to be $2\pi$ periodic in $\theta$ there must be $N$ different branches with the final form of the superpotential on the k$^{th}$ branch given by
\begin{equation}
W_{SYM} = c\, \omega_k \Lambda^3
\end{equation}
where $\omega_k = e^{k \frac{2\pi i}{N}}$. The full dynamics leading to this superpotential is that of gluino condensation, and leads to $c=N$, as was argued in~\cite{Affleck:1984xz,Cordes:1985um,Shifman:1987ia,Finnell:1995dr}. Since supersymmetry is unbroken, the potential is vanishing in the vacua of all of these branches, however the value of the gluino condensate does differ on each of them. 

Let us now turn on supersymmetry breaking in the form of AMSB. Since the SUSY breaking parameter $m$ appears as the F-term of the conformal compensator $\Phi = 1 + \vartheta^2 m$, $m$ has to have R-charge $-2$. Hence from $U(1)_R$ spurion analysis as well as the SUSY selection rules (i.e. Eq.~\eqref{eq:VtreeAMSB}) which include terms $\propto m W$ we expect a term in the potential\footnote{Note that AMSB allows two set of tree-level terms with a non-trivial $U(1)_R$ phase, $V_{\rm tree}\sim m W$ and $V_{\rm tree}\sim m\left(\partial_i W g^{i j^*} \partial_j^* K\right)$, but under our assumptions they have the same dependence on $\theta$, because $K$ and all the low energy fields are neutral under the $U(1)_R$. This remains true when we add $F\neq 0$ below.}
\be
V \propto m e^{i \theta/N}\, , 
\ee
which is indeed what we find in our full analysis (see for example Eq.~\eqref{eq:SQCDF0} in the next section). Again requiring that  physics be $2\pi$-periodic in $\theta$, we can conclude that the vacuum energy must have $N$ branches as a function of $\theta$.

 Let us now add $F$ flavors $Q, \bar Q$ in the fundamental and anti-fundamental representation of $SU(N)$ with vanishing R-charges. This will imply that the fermionic components have R-charge $-1$,   modifying the
 anomaly: the spurious charge of $\theta$ will now be $2(N-F)$\footnote{The Dynkin index of the fundamental representation is normalized to $1/2$.}
\be
\theta \to \theta + 2(N-F)\alpha\, .
\ee
 To see how this leads to the $\theta$ dependence of the potential and the branched structure, we introduce SUSY breaking and quark masses. We begin by considering the supersymmetric limit and the addition of quark masses $m_Q$. $m_Q$, as a spurion, has to carry R-charge $2$ (as well as being in the bifundamental representation of the flavor group). Assuming that the only light fields in the IR effective theory are supermesons or superbaryons which do not carry any R-charge (since $Q, \bar Q$ have R-charge zero), we can again easily identify the possible $\theta$-dependence of these theories. This assumption corresponds to taking $m_Q \ll |\Lambda|$, as we do in the following to make contact with QCD.
 
 Since the only objects that transform non-trivially are the two spurions $\theta$ and $m_Q$, we can conclude that the form of the superpotential must be
\be
W_{\rm eff}=a e^{i\frac{\theta}{N-F}}+b m_Q\, ,
\ee
purely from the $U(1)_R$ selection rules and holomorphy. Thus there will be  a term of the form  
\be
V \supset (a b^*) m_Q^* e^{\frac{i\theta}{N-F}}+{\rm h.c.}\, ,
\ee
in the scalar potential which together with the requirement of $2\pi$-periodicity in $\theta$ will imply the existence of $|N-F|$ branches. 

Finally let us add SUSY breaking via AMSB as we did for the pure SYM case. Following the same steps as above, we find two more terms allowed  in the scalar potential 
\be
V \supset b' (m m_Q) + a' (m e^{i \frac{\theta}{N-F}})+{\rm h.c.}\, ,
\ee
which are indeed both present in our explicit results below, see for example Eq.s~\eqref{eq:VUFlessN} and~\eqref{eq:VFgreaterN}. 
The $|N-F|$ branches are preserved by AMSB. It is interesting to notice that one can interpret the $|N-F|$ distinct vacua as a spontaneous breaking of the $Z_{2|N-F|}$ symmetry down to a $Z_2$, as pointed out in~\cite{Affleck:1983mk} for $F<N$.

The $\theta$-dependence of the potentials that we derive later in Sections~\ref{sec:FlessN} and \ref{sec:FlargerN} can be obtained purely from holomorphy and the $U(1)_R$ selection rules. However, a more precise construction requires to distinguish between $N>F$~\cite{Affleck:1983mk}, $N=F$~\cite{Seiberg:1994bz}, and $N<F$~\cite{Seiberg:1994pq}, as we do in Sections~\ref{sec:FlessN}, \ref{sec:FeqN} and \ref{sec:FlargerN}. In particular for $F=N$ the $U(1)_R$ is not anomalous and does not give us information about the $\theta$-dependence of the potential. We show in Section~\ref{sec:FeqN} how to analyze this case. 
However, we can still conclude that we expect a term proportional to $m m_Q$ (following the same arguments as for $F\neq N$) and that terms of the form $\Lambda^n$, which were forbidden for $F\neq N$, are instead allowed. We find both classes of terms in Eq.~\eqref{eq:VFeqN}. Note that the $m^2 \Lambda^2$ term in the same equation should be interpreted as a $|m|^2 \Lambda^2$ term, because in later sections we rotate away the phase of $m$ by shifting the $\eta'$ and $\theta$. Everywhere below we treat $m$ as a real parameter. This explains the form of Eq.s~\eqref{eq:PotentialFltN}, \eqref{eq:VFeqN}, \eqref{eq:VFN1} and~\eqref{eq:VFgreaterN}.

\section{$F<N$: the ADS superpotential. Modified branch structure and instanton generated $\eta'$ mass}\label{sec:FlessN}

We have already argued above what the expected form of $\theta$-dependence will be in SUSY QCD with AMSB. In the next three sections we present the full detailed evaluation of the chiral Lagrangian in these theories, which will also yield the explicit form of the $\theta$-dependence for each of these cases. 

Let us first investigate the simplest case of $F<N$. This was also explored in~\cite{Dine:2016sgq} by adding a mass to the squarks and gluinos, reaching results similar to those obtained here via the AMSB method. In the following we always assume $N>2$ so that the chiral symmetry and its breaking pattern can reflect those of ordinary QCD.

For $F < N$ the superpotential can be written in terms of the meson matrix $M_{ff'} = \bar{Q}_f Q_{f'}$ as
\begin{equation}
W = (N - F) \left( \frac{\Lambda^{3N-F}}{\det M}\right)^{1/(N-F)} + \text{Tr} (m_Q M)\,,
\end{equation}
where the first term is the non-perturbative ADS superpotential~\cite{Davis:1983mz,Affleck:1983mk} and the second is a mass term for the quark superfields. Here $\Lambda$ is the holomorphic scale, which is scale invariant\footnote{For the conversion between the holomorphic scale and the physical scale see Appendix~\ref{app:holomScale}.}. Note that $m_Q$ can always be diagonalized by a bi-unitary transformation. In the following we will assume the hierarchy $\text{Tr}\, m_Q \ll m \ll \Lambda$, i.e. $m_Q$ is a small spurion that explicitly breaks the $U(F)\times U(F)$ flavor symmetry, just like the quark masses in  QCD.

We parameterize the $D$-flat directions as $Q^a_f = \bar{Q}^a_f = f \delta^a_f$ which implies $M_{f f'}= f^2 \delta_{f f'}$ and determine the scalar potential for $f$ using Eq.~\eqref{eq:VtreeAMSB}\footnote{Note that the K\"ahler potential for $f$ is not canonical $K = 2F f^\dagger f$ and consequently $g_{f f^\dagger} = 2F$.}
\be
V &=& (2 F)^{-1} \left| \frac{2F}{f} \left( \frac{\Lambda^{3N - F}}{f^{2F}}\right)^{1/(N-F)} - 2 f \text{Tr}( m_Q) \right|^2 \nn \\
&-& m\left[(3N-F)  \left( \frac{\Lambda^{3N - F}}{f^{2F}}\right)^{1/(N-F)} + f^2 \text{Tr}(m_Q)   \right] + {\rm h.c.} 
\label{eq:FsmallerNPot}
\ee
We neglect the contributions to the potential from the SUSY breaking squark mass terms (see Eq.~\eqref{susybreakingmasses}), since they are loop suppressed and give a negligible contribution to the minimization of the potential. Like the D-terms, these terms also do not depend on the Goldstone bosons.

\noindent For $\text{Tr}\, m_Q \ll m\ll \Lambda$ the scalar potential in Eq.~\eqref{eq:FsmallerNPot} is minimized for
\be
|f| &=& |\Lambda| \left( \frac{N+F}{3N-F}\frac{|\Lambda|}{m}\right)^{(N-F)/(2N)} + \mathcal{O}(m_Q / m)\,. 
\label{eq:fVEV}
\ee
Thus the $U(F)\times U(F)$ flavor symmetry is spontaneously broken to its diagonal subgroup $U(F)_V$. 
 The GBs of the $U(F)\times U(F) \to U(F)_V$ breaking are parameterized by a unitary matrix $U$ (cf~\cite{Dine:2016sgq})
\begin{equation} \label{eq:UParam}
Q^a_f = |f | \delta^a_f\,,\quad \bar{Q}^a_f = Q^a_{f'} U_{f' f}\,,\quad M = |f |^2 U\,.
\end{equation}

The resulting  scalar potential for $U$ obtained from the potential for $Q$ and $\bar{Q}$ after the substitution of Eq.~\eqref{eq:UParam} is
\begin{equation}
\label{eq:VUFlessN}
\begin{split}
V = &-m\left[ (3N -F)  \left(\frac{\Lambda^{3N - F}}{|f |^{2F}}\right)^{1/(N-F)} \det (U)^{-1/(N-F)} +  |f |^2 \text{Tr}( m_Q U)\right] + {\rm h.c.}\\
&-2\left(\frac{\Lambda^{3N - F}}{|f |^{2F}}\right)^{1/(N-F)} \det (U)^{-1/(N-F)} \text{Tr}(m_Q^\dagger U^\dagger) +{\rm h.c}\,.
\end{split}
\end{equation}
As in the previous sections, we are only interested in the dependence on $\eta'$, $\theta$ and the remaining neutral GBs, so we take $U= e^{i \eta'} e^{i \pi^j t^j}$, where $t^j$ are the generators of the Cartan sub-algebra of $SU(F)$. We would also like to remind the reader that $\eta'$ in our normalization has a periodicity of $\tfrac{2\pi}{F}$. Again, the fields are taken to be dimensionless: the physical canonically normalized fields can be obtained with the replacement $\eta' \to \eta'/(\sqrt{2F} f)$ and $\pi \to \pi/ (2f)$, with the identification $f_\pi = f_{\eta'} = f$. 

In terms of the $\eta'$ and pions the scalar potential reads $V=\min_k V_k$, with
\begin{equation}\label{eq:PotentialFltN}
\begin{split}
V_k = &-2 (3N -F) \left(\frac{N+F}{3N - F}\right)^{-F/N} \left(\frac{m}{|\Lambda |}\right)^{F/N} m |\Lambda |^3 \cos\left( \frac{F}{N-F}\eta' - \frac{\theta + 2\pi k}{N-F} \right)\\
&-2  \left(\frac{N+F}{3N - F}\right)^{1-F/N} \left(\frac{m}{|\Lambda |}\right)^{F/N} |\Lambda |^3 \sum_{i=1}^F m_i \cos \left( \eta' +\theta_Q + \sum_{j=1}^{F-1} t^j_i \pi_j \right)\\
&-4 \left(\frac{N+F}{3N - F}\right)^{-F/N} \left(\frac{m}{|\Lambda |}\right)^{F/N}  |\Lambda |^3  \sum_{i=1}^{F} m_i \cos\left( \frac{N}{N-F} \eta' + \theta_Q - \frac{\theta + 2\pi k}{N-F} + \sum_{j=1}^{F-1} t^j_i \pi_j\right)\,,
\end{split}
\end{equation}
where $\Lambda^{3N-F} = |\Lambda |^{3N-F} e^{i\theta}$. Here  $m_i$ is the i$^{th}$ diagonal element of the quark mass matrix, ${\theta_Q = \arg\det m_Q}$ is the phase of the mass matrix, $t^j_i$ is the i$^{th}$ diagonal element of the j$^{th}$ Cartan generator. The factors $2\pi k/(N-F)$ in the cosine comes from the branches of the complex root of the ADS superpotential.

While there is an unmistakable similarity between the first term in Eq.~\eqref{eq:PotentialFltN} and Eq.~\eqref{eq:QCDactual} and between the second and third term with Eq.~\eqref{eq:QCDGBMassTerm} in the QCD chiral Lagrangian, there are some important qualitative differences. As expected, the potential exhibits a branch-like structure   due to the non-analyticity induced by gaugino condensation.     In contrast to pure QCD the number of branches is not $N$ but $N-F$. This is the consequence of the dynamics of the SUSY theory: with $F$ flavors there are also $F$ squarks that break the gauge group to $SU(N-F)$.  Then gaugino condensation in this unbroken group gives rise to the $N-F$ branches. The most important lesson  here is that the introduction of flavors does actually change the dynamics of confinement: instead of the $N$ branches there are only $N-F$ branches, and the assumption that the potential of the theory with flavors is simply the potential of the confining theory with the replacement $\theta \to \theta - F \eta'$ does not hold in this case.

Assuming that the first term in Eq.~\eqref{eq:PotentialFltN} dominates the $\eta'$ potential, which is the case for $\text{Tr}\, m_Q / m \ll F^2/N$,  a simple analytic expression for the $\eta'$ mass is found to be
\begin{equation}
    m_{\eta'}^2 = \frac{ (x-3)^2 x}{(x+1)(x-1)^2}m^2\,,\quad \text{with}\quad x=\frac{F}{N}\,.
\end{equation}
The mass of the $\eta '$ scales as $m_{\eta '} \propto 1/N$ in the $N\gg F$ limit as predicted by the Veneziano-Witten formula. This expression depends only on the ratio $x=F/N$, i.e. it is finite in the large $N$ limit if also the number of flavors is large with a fixed ratio $F/N$. This is not too surprising: the anomaly equation with $F$ flavors Eq.~(\ref{eq:anomalyF}) shows that if $F\propto N$, then the anomaly does not vanish in the large $N$ limit, and there is no reason to expect the $\eta'$ mass to vanish. 
The mass is a monotonously growing function for $0\leq x < 1$ and diverges at $x=1$ where also our current treatment of SQCD breaks down. The pole at $x=1$ is related to the breakdown of the effective Lagrangian in Eq.~(\ref{eq:PotentialFltN}) for large $N$ in the limit where $N-F$ is held fixed.

Under the assumption that $\text{Tr}\, m_Q / m \ll F^2/N$ it is straightforward to integrate out the $\eta'$ as the first term dominates and fixes
\begin{equation}
\eta' = \frac{\theta + 2\pi k}{F} + \frac{N-F}{F} 2\pi j\,,
\end{equation}
where $j$ labels the infinite set of solutions due to the periodicity of the cosine. Restricting $\eta'\in [0,\tfrac{2\pi}{F}]$ there is a unique minimum $\eta' = \tfrac{\theta}{F}$, corresponding to $k=j=0$. The remaining terms 
give the following potential for the neutral GBs and $\theta$ 
\begin{equation}\label{eq:AMSBPotential}
V =  -2\frac{7N-F}{3N-F}\left(\frac{N+F}{3N - F}\right)^{-F/N} \left(\frac{m}{|\Lambda |}\right)^{F/N} |\Lambda |^3 \sum_{i=1}^{F} m_i \cos\left( \frac{\theta + F \, \theta_Q}{F} + \sum_{j=1}^{F-1} t^j_i \pi^j \right)\,.
\end{equation}

Despite the different branch structure of the $\eta '$ potential in Eq.~\eqref{eq:PotentialFltN} compared to the chiral Lagrangian in QCD, after integrating out the $\eta'$ we arrive at a potential which has exactly the same structure as Eq.~\eqref{eq:QCDGBPotential}. The reason is   that if the $\eta '$ is heavier than the remaining GBs it completely washes out any branch structure of the original potential---it acts as a heavy QCD scale axion. This is in agreement with the result in~\cite{Witten:1980sp} where it is shown that for $m_u m_d< m_s|m_d-m_u|$ in large $N$ QCD with three flavors, the pion potential does not have branches.

In order to find the potential for $\theta$ the GBs $\pi^j$ need to be integrated out. Without an explicit additional light axion this is more complicated than in Section~\ref{sec:axionMass}, since an analytic solution for the minimization conditions is not known for general $F$. However, the solution for $F=2$ and $F=3$ are analogous to the QCD chiral Lagrangian and lead to a smooth 
and $2\pi$-periodic vacuum energy for non-degenerate masses $m_i \neq m_j$. Degenerate masses, on the other hand, cause cusp-like features when at branch-transitions.

It is instructive to consider a few special cases for the scalar potential in Eq.~\eqref{eq:PotentialFltN}. The simplest case is pure SYM theory with $F=0$.\footnote{While the ADS superpotential is not well-defined for $F=0$ this naive extrapolation yields the same result as the computation with the low-energy effective superpotential for gaugino condensation $W=N\,\Lambda^3$ in AMSB.} In this scenario the contribution to the scalar potential comes purely from gaugino condensation and has the form
\begin{equation}\label{eq:SQCDF0}
V_k\xrightarrow{F=0} -6 N^2 m |\Lambda_{\rm phys} |^3 \cos\left( \frac{\theta + 2\pi k}{N} \right)\,,
\end{equation}
where we used that in pure SYM $|\Lambda | = N^{1/3} |\Lambda_{\rm phys}|$ (see Appendix~\ref{app:holomScale}). This result reproduces the expectation from gluodynamics in QCD that the vacuum energy is a function of the form $N^2 f(\theta/N)$ with $N$ branches.

Adding a small number of flavors, i.e. taking the $F\ll N$ limit, the potential simplifies to
\begin{equation}
V_k \stackrel{N\gg F}{\rightarrow} -6N^2 m |\Lambda_{\rm phys} |^3 \cos \left( \frac{F}{N} \eta' - \frac{\theta + 2\pi k}{N}\right) -\frac{14}{3} N  |\Lambda_{\rm phys}|^3 \sum_{i=1}^{F} m_i \cos\!\left( \eta' + \theta_Q + \sum_{j=1}^{F-1} t^j_i \pi^j\right).
\end{equation}
 This nicely shows that the leading term in the large $N$ limit still comes from gaugino condensation in the unbroken part of the group, whereas quark contributions are suppressed by one power of $N$. It is also straightforward to see that for $m_Q = 0$, i.e. when the axial symmetry at the classical level is unbroken, the $\eta'$ mass $m_{\eta'}^2 \propto F m |\Lambda_{\rm phys}|^3 / f^2 \sim F/N$ vanishes in the $N\rightarrow \infty$ limit ($f \sim \sqrt{N}$) and the $\eta'$ becomes an exact GB, which is a consequence of the anomaly term vanishing in the large $N$ limit. In this limit $\theta$ is unphysical as it can be absorbed in the definition of $\eta'$. 

The situation changes when both $F$ and $N$ are large. For $F = N-1$ the ADS superpotential is generated by instantons and the branched structure for the combined $\{\eta',\theta\}$ potential completely disappears, i.e. it is automatically $2\pi$ periodic in $\theta$. This is an example where the dynamics produces a potential without branches, and the mechanism is due to a calculable 1-instanton effect. This will persist both for finite and for large $N$, as long as $F=N-1$. 
In the limit $N\gg 1$, with $F=N - 1$ fixed, the potential takes the form
\begin{equation}\label{eq:FisNm1}
\begin{split}
V_k \stackrel{N=F+1 \gg 1}{\rightarrow} &-4 N^{3/2} m^2 |\Lambda_{\rm phys} |^2 \cos\left( (N-1) \eta' - \theta\right) \\
&- 2 N^{1/2} m |\Lambda_{\rm phys}|^2 \sum_{i=1}^{F} m_i\cos\left(\eta' + \theta_Q  + \sum_{j=1}^{F-1} t^j_i \pi^j\right) \\
&-4 N^{1/2} m |\Lambda_{\rm phys}|^2  \sum_{i=1}^{F} m_i\cos\left(N \eta' + \theta_Q -\theta   + \sum_{j=1}^{F-1} t^j_i \pi^j \right)\,.
\end{split}
\end{equation}
The magnitude of this potential is set by $m^2 \Lambda^2$, and is not vanishing in the large $N$ limit. While the instanton action is proportional to  $\Lambda^{2N+1} \propto e^{-N}$, the potential still  remains finite at large $N$.

 Another striking feature is that all terms have the same scaling with $N$ with a non-integer exponent if all masses are degenerate ($\sum_i m_q = F m_q$).  Note however that this Lagrangian becomes strongly coupled in the large $N$ limit, just like the more general Eq.~(\ref{eq:PotentialFltN}) for the limit where $N-F=p$ is held fixed (rather than the $x=F/N$ ratio).

%
\subsection{Vacuum structure and phase transition}
%
Before moving on to $F\geq N$ in the next section, we will comment on the vacuum structure and compare it to results obtained for QCD using large $N$ methods~\cite{Witten:1978bc,Witten:1979vv,Witten:1980sp,DiVecchia:1980yfw,Witten:1998uka} and arguments based on anomalies~\cite{Gaiotto:2017yup,Gaiotto:2017tne} for finite $N$.

For $F=0$ it has been shown in the large $N$ limit that the theory possesses a unique vacuum for generic values of $\theta$ and undergoes a first-order phase transition as $\theta$ is moved through $\pi$. This happens since in large $N$ QCD the vacuum energy is branched and non-analytic at points where the branches cross (see Section~\ref{sec:largeN}). This means that at $\theta=\pi$ a jump between 
 two degenerate vacua occurs and CP is spontaneously broken. 
 In the supersymmetric version the vacuum energy for $F=0$ in Eq.~\eqref{eq:SQCDF0} has the same structure as in large $N$ QCD and therefore has a doubly-degenerate vacuum at $\theta=\pi$ which means CP is spontaneously broken at this point.

In~\cite{Gaiotto:2017tne} it was argued that for $F=1$ at $\theta = \pi$ the theory has two degenerate vacua and therefore spontaneously breaks CP, but only for large quark masses $m_Q$ above a critical value, i.e. $|m_Q| > |m_{Q,0}|$. Below the critical value there is always a unique vacuum and CP is not spontaneously broken. At the critical value $\eta'$ becomes exactly massless. Since we will find that the critical value is at masses of the order $m_{Q,0}\sim {m}/N$, we can only reliably observe the transition between these two regimes in the large $N$ limit. 

The scalar potential for $F=1$ has the form
\begin{equation}
    V_k(\eta',\bar{\theta}) \propto -a\, m |\Lambda|^3\cos\left(\frac{\eta' - (\bar{\theta}+2\pi k)}{N-1}\right) - b\, m_Q |\Lambda|^3\cos(\eta') - 2 m_Q |\Lambda|^3 \cos\left(\frac{N \eta' - (\bar{\theta}+2\pi k)}{N-1}\right)\,, \label{Feq1}
\end{equation}
where $a=(3N-1)$, $b=\frac{N+1}{3N-1}$ and we dropped a global factor of $b^{-1/N}(m/|\Lambda |)^{1/N}$. For $m\gg m_Q$ the first term dominates and fixes the $\eta'$ VEV and we can choose $k=0$ and thus $\eta' = \bar{\theta}$ as a minimum.\footnote{Recall that $\eta'$ is an angular variable and is defined modulo $2\pi$ for $F=1$ (and $\tfrac{2\pi}{F}$ in general). Thus if we want to restrict it to values in the interval $[0,2\pi ]$ this fixes $k$.} Plugging this into the last two terms we see that the potential is regular at $\bar{\theta}=\pi$
\begin{equation}
    V(\bar{\theta}) \propto  - (b+2)\, m_Q |\Lambda|^3\cos(\bar{\theta}) \,,
\end{equation}
and no phase transition occurs.

\begin{figure}[t!]
    \centering
    \subfigure{\includegraphics[width=0.33\textwidth]{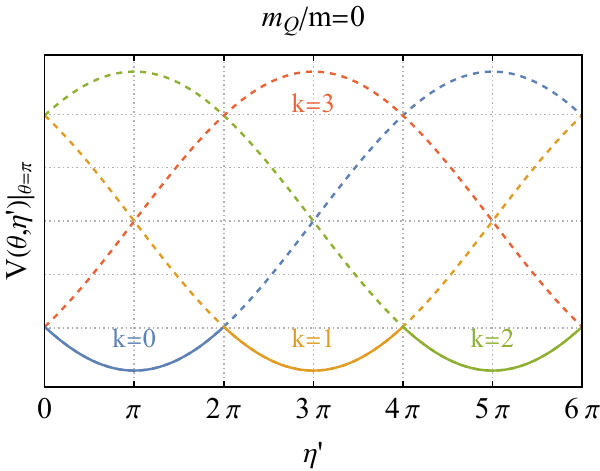}}
    \hspace{-.2 cm}
    \subfigure{\includegraphics[width=0.33\textwidth]{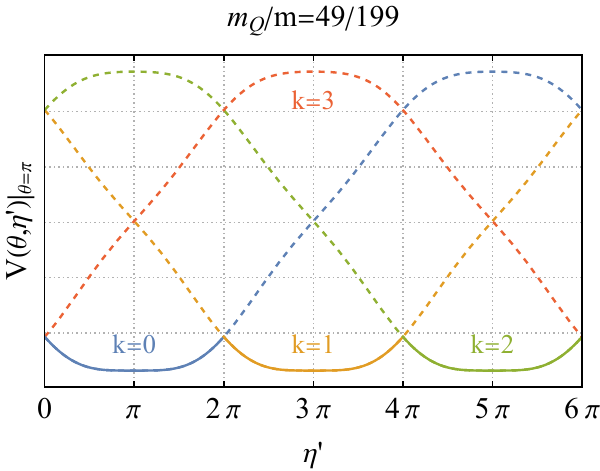}}
        \hspace{-.2 cm}
\subfigure{\includegraphics[width=0.33\textwidth]{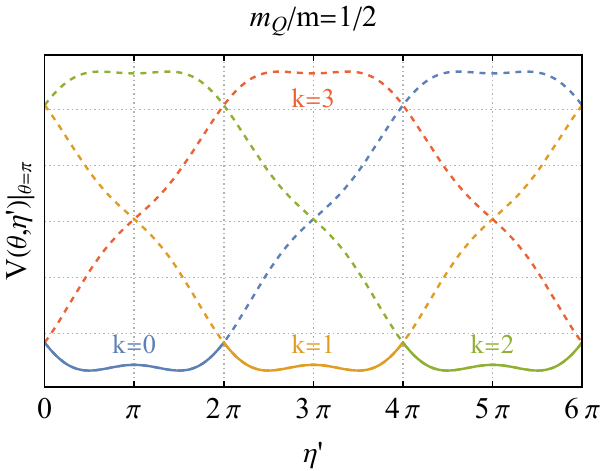}}
    
    \caption{The  $\eta'$ potential according to Eq.~\eqref{Feq1} for $N=5$ and  $F=1$ along $\theta=\pi$. The four branches are depicted in different colors. The actual potential is the lower envelope of the branches, i.e. the solid curve, which is $2\pi$-periodic. Note that the physical field range for the $\eta'$ is $\eta'\in [0,2\pi ]$ for which the potential reduces to the $k=0$ branch. For $m_Q < m_{Q,0}  $ (left), the minimum of the potential occurs for $\eta'=\theta$ and CP is conserved.  For $m_Q = m_{Q,0} =(49/199)m$ (middle), the potential is flat and the $\eta'$ is massless. Above the critical point $m_Q > m_{Q,0} $, a phase transition occurs and the minimum of the potential moves away from the CP conserving minimum and CP is spontaneously broken.}
    \label{fig:phasetransition}
\end{figure}

Taking the second derivate of the potential Eq.~\eqref{Feq1} to find the mass of the $\eta'$, it is straightforward to show that the $\eta'$ becomes massless for 
\begin{equation}
m_Q = m_{Q,0} = \frac{(3N-1)^2}{7 N^3-3 N^2-N+1} m\,.
\end{equation}
Therefore we can only reliably trust the results for large $N$ where $m_{Q,0} \ll m$ which we assumed in the minimization of the potential. For $m_Q > m_{Q,0}$ there are two degenerate minima which we show in Fig.~\ref{fig:phasetransition} for $N=5$, where the critical mass is $m_{Q,0} =49/199$.

The situation is different for $F>1$ with equal quark masses. In such a setup there are CP-conjugate degenerate vacua at $\bar{\theta}=\pi$ for all masses $|m_Q|>0$~\cite{Gaiotto:2017tne}. In particular this includes the region where $m_Q \ll m\ll \Lambda$ what we assume throughout the paper. In order to check if this is consistent with our results we follow~\cite{Gaiotto:2017tne} and take equal masses for all matter fields, i.e. $m_Q = m_Q \delta_{ff'}$, and assume that the pion VEVs do not break the residual $SU(F)$ flavor symmetry.\footnote{In~\cite{Gaiotto:2017tne} it was shown that the $SU(F)$ preserving solutions are true local minima.} If this is the case the VEV of $e^{i\pi^a T^a}$ has to be in the center of $SU(F)$, i.e. $U$ has to be of the form
\begin{equation}
    U= e^{i\eta'} e^{2\pi i l/F} \mathbb{1}\,,
\end{equation}
where $l=0,1,\ldots , F-1$. The resulting potential is obtained from Eq.~\eqref{eq:PotentialFltN} after the replacement $\sum_j \pi^j t^j_i \rightarrow 2\pi l/F$ and $m_i \rightarrow m_Q$. The first term again dominates for $m \gg m_Q$ and fixes $\eta' = \bar{\theta}/F$ where we restricted $\eta'$ to its physical field range $\eta'\in [0,\tfrac{2\pi}{F}]$.
Plugging this into the remaining terms yields
\begin{equation}
    V_{l}(\bar{\theta}) \propto - F m_Q |\Lambda|^3 \cos\left(\frac{\bar{\theta}}{F} + \frac{2\pi l}{F}\right)\,.
\end{equation}
For generic values of $\bar{\theta}$ this has a unique minimum. E.g. for $\bar{\theta}=0$ the potential is clearly minimized for $l=0$. For $\bar{\theta}=\pi$ on the other hand the potential takes the form
\begin{equation}
    V_l(\pi) \propto - F m_Q |\Lambda|^3 \cos\left(\frac{(2l+1)\pi }{F} \right)\,,
\end{equation}
which has two minima: $l=0$ and $l=F-1$. Thus for $\bar{\theta}=\pi$ the vacuum configurations are $U= e^{\pm i\pi/F} \mathbb{1}$. These are related by the CP transformation $U\rightarrow U^\dagger$ which implies that CP is spontaneously broken irrespective of the size of $m_Q$.
%
\section{$F=N,N+1$: the confining cases}\label{sec:FeqN}
For $F=N$ all the ‘t Hooft anomaly matching conditions can be solved in a confining theory with color singlet degrees of freedom~\cite{Seiberg:1994bz}: the meson matrix $M_{f f'}$ and the baryon fields $B= \epsilon ^{f_1\cdots f_N} B_{f_1\cdots f_N}$ and $\bar{B} = \epsilon ^{f_1\cdots f_N} \bar{B}_{f_1\cdots f_N}$, where $B_{f_1\cdots f_N}$ and $\bar{B}_{f_1\cdots f_N}$ are the completely antisymmetric color singlet combinations of the quark and anti-quark superfields $Q$ and $\bar{Q}$, respectively. The degrees of freedom describing the moduli space satisfy a quantum modified constraint
\begin{equation}\label{eq:quantConstraint}
\det(M) - \bar{B} B = \Lambda^{2N}\,.
\end{equation}
This constraint is implemented in the superpotential with the help of a Lagrange multiplier superfield $X$
\begin{equation}
W = X\left( \frac{\det (M) - \bar{B} B}{\Lambda^{2N}} - 1 \right) +  \text{Tr}(m_Q M)\,.
\end{equation}
Note that we chose to implement the constraint on $\det(M)$ and $\bar{B}B$ such that $X$ does not carry a charge under the spurious $U(1)_A$ axial symmetry. Interpreting $X$ as a dynamical degree of freedom we consider the K\"ahler potential
\begin{equation}
K = \frac{\text{Tr}(M^\dagger M)}{\alpha |\Lambda|^2}+ \frac{X^\dagger X}{\beta |\Lambda|^4} + \frac{\bar{B}^\dagger \bar{B}}{\gamma |\Lambda |^{2N -2}} + \frac{B^\dagger B}{\delta |\Lambda |^{2N-2}}\,,
\end{equation}
where $\alpha,\beta,\gamma,\delta$ are unknown $\mathcal{O}(1)$ numbers. Note that keeping only the quadratic terms in the K\"ahler potential is justified if $M, X,B,\bar{B} \ll \Lambda$, which will turn out not to be the case. A more solid approach is to start from $F=N+1$ and then give one flavor a heavy supersymmetric mass $\mu$ with $\Lambda \gg \mu \gg m \gg \text{Tr}\, m_Q$ and integrate it out. In this approach the Lagrange multiplier field $X$ will be identified with the $M_{N+1,N+1}$ component of the meson field, justifying the assumption on its Kahler potential above.  We have checked that this procedure gives results which are compatible with our simplified approach. To leading order in $m$ and $m_Q$ the resulting scalar potential has a minimum for
\begin{equation}
M_{ff'} = f^2 \delta_{ff'}\,,\quad X = -\frac{m|\Lambda|^2}{\alpha}\,,\quad B= \bar{B} =0\,,\quad \text{with}\quad f=\Lambda\,.
\end{equation}
Whether this is the global minimum depends on the values of the unknown $\mathcal{O}(1)$ numbers $\alpha$ and $\beta$~\cite{Murayama:2021xfj,Csaki:2022cyg}. In the following we will assume that this chiral symmetry breaking minimum is the global minimum.
Parameterizing again the neutral GBs as $M = |f|^2 U$ with $U= e^{i \eta'} e^{i \pi^a t^a}$, where the decay constant $f$ is absorbed into the pion an $\eta'$ fields, we find a potential that is given by
\begin{equation}
\begin{split}
V = &-2 |\Lambda |^4  \left[\beta + (N-2) \frac{m^2}{\alpha |\Lambda|^2}\right] \cos\left( N \eta' - \theta \right) -4 m |\Lambda |^2 \sum_{i=1}^N m_i \cos\left( \eta' + \theta_Q + \sum_{j=1}^{N-1} t^j_i \pi^j \right) \\
& - 2 m |\Lambda |^2 \sum_{i=1}^N m_i \cos\left( (N-1) \eta' - \theta_Q - \theta -\sum_{j=1}^{N-1} t^j_i \pi^j \right)\,.
\label{eq:VFeqN}
\end{split}
\end{equation}
The structure of the potential, including the scaling of the prefactors, is very similar to Eq.~\eqref{eq:FisNm1}. In particular there is no branch-like structure and the potential is a pure one-instanton effect, i.e. it is proportional to $e^{\pm i\theta}$. One difference is, however, that the first term is enhanced, i.e. the pure $\eta'$ potential here scales as $|\Lambda|^4$ instead of $m^2 |\Lambda|^2$ in Eq.~\eqref{eq:FisNm1}.\footnote{In the alternative derivation where we integrate out one flavor from the $F=N+1$ case the scaling is $|\Lambda|^2\mu^2$ which is still much larger than $|\Lambda|^2 m^2$.} Consequently the leading contribution to the $\eta'$ mass is proportional to $\Lambda$ instead of the SUSY breaking scale $m$, as it was the case for $F<N$. This is a direct consequence of the quantum modified constraint on the moduli space in Eq.~\eqref{eq:quantConstraint} which breaks the axial symmetry already before SUSY breaking is introduced.
Note that just as in ordinary QCD when $m_Q=m=0$ physical quantities do not depend on $\theta$. The dependence of the $|\Lambda|^4$ term can be shifted away by a redefinition of the $\eta^\prime$ field that leaves the rest of the action invariant.
Integrating out the $\eta'$ gives
\begin{equation}
V_k = -6 m |\Lambda |^2 \sum_{i=1}^N m_i \cos\left( \frac{\theta + N \, \theta_Q + 2\pi k}{N}  + \sum_{j=1}^{N-1} t^j_i \pi^j \right)\,,
\end{equation}
which is a straightforward extrapolation of the $F=N-1$ case.

For $F=N+1$ the baryons $B^f = \epsilon^{f_1\cdots f_N f} B_{f_1\cdots f_N}$ and antibaryons $\bar{B}_f = \epsilon_{f_1\cdots f_N f}\bar{B}^{f_1\cdots f_N}$ transform in the antifundamental and fundamental representation of $SU(F)$, respectively. In this case the classical and quantum constraints are identical and follow from the superpotential
\begin{equation}
W = \frac{B M \bar{B} - \det (M)}{\Lambda^{2N-1}} + \text{Tr}(m_Q M)\,,
\end{equation}
where the contraction of flavor indices is implicit in the first term and we also added a mass term. The K\"ahler potential up to quadratic order in the fields is of the form
\begin{equation}
K = \frac{\text{Tr}(M^\dagger M)}{\alpha |\Lambda|^2} + \sum_f \frac{\bar{B}_f^\dagger \bar{B}_f}{\beta |\Lambda |^{2N -2}} + \sum_f \frac{B_f^\dagger B_f}{\gamma |\Lambda |^{2N-2}}\,,
\end{equation}
where $\alpha,\beta,\gamma$ are unknown $\mathcal{O}(1)$ numbers, which for simplicity we will set to one in the following. The corresponding scalar potential is minimized for
\begin{equation}
M_{ff'} = f^2 \delta_{ff'}\,,\quad B_f = \bar{B}_f =0\,,\quad \text{with}\quad |f|^2 = |\Lambda|^2 \left(\frac{N-2}{N} \frac{m}{|\Lambda|}\right)^{1/(N-1)}\,.
\end{equation}
Note that in contrast to the $N=F$ case $|f| \ll |\Lambda |$ which justifies keeping only terms quadratic in the fields in the K\"ahler potential.

We again introduce the GBs as $M = |f|^2 U$ (including the phase of $f$ is just a shift in the definition of the $\eta^\prime$), resulting in the following potential
\begin{equation}
\begin{split}\label{eq:VFN1}
V = &-2 (N-2) \left(\frac{N-2}{N}\frac{m}{|\Lambda|}\right)^{(N+1)/(N-1)} m |\Lambda|^3 \cos\left( (N+1) \eta' - \theta\right)\\
&-2 \left(\frac{N-2}{N}\frac{m}{|\Lambda|}\right)^{N/(N-1)} |\Lambda|^3 \sum_{i=1}^{N+1} m_i  \cos\left( N \eta'- \theta_Q - \theta - \sum_{j=1}^{N} t^j_i \pi^j\right)\\
&-4  \left(\frac{N-2}{N}\frac{m}{|\Lambda|}\right)^{1/(N-1)} m  |\Lambda|^2 \sum_{i=1}^{N+1} m_i \cos\left( \eta'+ \theta_Q + \sum_{j=1}^{N} t^j_i \pi^j\right)\,.
\end{split}
\end{equation}
Similarly to the $F=N$ and $F=N-1$ cases also this potential does not have a branch-like structure and is consistent with an instanton effect.
After integrating out the $\eta'$ we again obtain a potential for the pions which has the same structure
\begin{equation}
V = -2\frac{3N-2}{N} \left(\frac{N-2}{N}\frac{m}{|\Lambda|}\right)^{1/(N-1)} m |\Lambda |^2 \sum_{i=1}^{N+1} m_i \cos\left( \frac{\theta + (N+1) \, \theta_Q}{N+1} + \sum_{j=1}^{N} t^j_i \pi^j\right)\,.
\end{equation}
In the large $N$ limit $|\Lambda| = N^{1/4} |\Lambda_{\rm phys}|$, and the potential energy scales as $V\propto N^{3/2}$ for $\sum_i m_i \propto N$.

\section{$F>N+1$: Gaugino condensation in the dual gauge group}\label{sec:FlargerN}

For $N+1 < F < 3/2 N$ we can study the low-energy dynamics in a weakly coupled dual $SU(F-N)$ gauge theory with dynamical scale $\tilde{\Lambda}$. This theory contains $F$ (anti-)fundamentals $q$ $(\bar{q})$ under $SU(F-N)$ and the meson matrix $M$, which we identify with the meson matrix that appears in the original theory. The superpotential is given by
\begin{equation}
W_d = \frac{1}{\mu} q_i M_{ij} \bar{q}_j +  \text{Tr}(m_Q M)\,,
\end{equation}
where $\mu$ is a scale which appears in the relation between the dynamical scales of the original and dual theories
\begin{equation}\label{eq:scaleRelation}
\Lambda^{3N-F} \tilde{\Lambda}^{3\tilde{N}-F} = (-1)^{F-N}\mu^F\,,
\end{equation}
where we introduced $\tilde{N}= F-N$ to make the relation more symmetric. We consider a K\"ahler potential which contains the dynamical scale of the original theory~\cite{Intriligator:2006dd}
\begin{equation}
K = \frac{\text{Tr}(M^\dagger M)}{|\Lambda|^2} +  \sum_i q_i^\dagger e^{\tilde{V}} q_i + \sum_i \bar{q}_i^\dagger e^{\tilde{V}} \bar{q}_i \,,
\end{equation}
where we set unknown $\mathcal{O}(1)$ numbers to one. We assume that $M$ gets a VEV of the form $\langle M_{f f'}\rangle = f^2 \delta_{ff'}$, so that $SU(F)\times SU(F)\rightarrow SU(F)_V$. Plugging this in the superpotential generates a mass term for $q$ and $\bar{q}$ of size $\tilde{m}=f^2/\mu = \det (M/\mu)^{1/F}$. Thus in order to describe the low-energy dynamics we can integrate out $q$ and $\bar{q}$, such that we have a pure $SU(F-N)$ gauge theory below $\tilde{m}$. The dynamical scale of the low-energy effective theory $\tilde{\Lambda}_{\rm eff}$ is obtained from the matching condition
\begin{equation}
    \left(\frac{\tilde{\Lambda}_{\rm eff}}{\tilde{m}}\right)^{3\tilde{N}}= \left( \frac{\tilde{\Lambda}}{\tilde{m}}\right)^{3\tilde{N}-F}\,.
\end{equation}
Gaugino condensation generates an effective superpotential for the low-energy effective theory
\begin{equation}
    W_d^{\rm eff} = \tilde{N} \tilde{\Lambda}_{\rm eff}^3 + \text{Tr}(m_Q M) = (N-F) \left(\frac{\Lambda^{3N-F}}{\det M}\right)^{\frac{1}{N-F}} + \text{Tr}(m_Q M)\,,
\end{equation}
which is the ADS superpotential after using Eq.~\eqref{eq:scaleRelation}. This implies that SQCD is symmetric around $F=N$, i.e. the dynamics governing the $F>N+1$ scenario is secretly the same as for the $F<N$ case. The potential is generated by gaugino condensation in the unbroken part of the gauge or dual gauge group, respectively.
Now we can simply compute the scalar potential with Eq.~\eqref{eq:VtreeAMSB}. The minimum to leading order in $m_Q$ is given by
\begin{equation}
M_{ff'} = f^2 \delta_{ff^\prime}\,,\quad \text{with} \quad |f|^2 = |\Lambda|^2 \left( \frac{3N-2F}{N} \frac{m}{|\Lambda|} \right)^{\frac{F-N}{2N-F}}\,.
\end{equation}
With this it is straightforward to find the potential for the neutral GBs
\begin{equation}
\begin{split}
V=& \min_k V_k \label{eq:VFgreaterN} \\
V_k = &-2(3N-2F) \left( \frac{3N-2F}{N}\frac{m}{|\Lambda|}\right)^{F/(2N-F)} m |\Lambda |^3 \cos\left( \frac{F}{F-N} \eta' - \frac{\theta + 2\pi k}{F-N} \right)\\
&-2 \left( \frac{3N-2F}{N}\frac{m}{|\Lambda|}\right)^{N/(2N-F)} |\Lambda |^3 \sum_{i=1}^{F} m_i \cos\left( \frac{N}{F-N} \eta' - \theta_Q - \frac{\theta + 2\pi k}{F-N} -\sum_{j=1}^{F-1} t^j_i \pi^j\right)\\
&-\frac{4 N}{3N-2F} \left( \frac{3N-2F}{N}\frac{m}{|\Lambda|}\right)^{N/(2N-F)}   |\Lambda |^3 \sum_{i=1}^{F} m_i \cos\left( \eta' + \theta_Q + \sum_{j=1}^{F-1} t^j_i \pi^j\right)\,,
\end{split}
\end{equation}
which has the same structure as Eq.~\eqref{eq:PotentialFltN} for $F<N$. However, there is a subtle difference. The number of branches in the $\eta'$ potential changed from $N-F$ to $F-N$ supporting that the structure of the potential is symmetric in $F$ around $F=N$. $F-N$ in this case appears since it is gaugino condensation in the dual gauge group which is responsible for generating the $\eta'$ potential.

Once we integrate out $\eta'$ we find again a potential of the following form
%
\begin{equation}
V =-\frac{2(5N-2F)}{3N-2F} \left( \frac{3N-2F}{N}\frac{m}{|\Lambda|}\right)^{N/(2N-F)} |\Lambda |^3 \sum_{i=1}^{F} m_i \cos\left( \frac{\theta  + F\,\theta_Q}{F} + \sum_{j=1}^{F-1} t^j_i \pi^j\right)\, .
\end{equation}
%
 
\section{Conclusions}\label{sec:conclusion}

We investigated the dynamics behind the potential of the $\eta'$, and consequently also the axion mass from QCD effects, in strongly coupled QCD-like theories. These models are based on ${\cal N}=1$ SUSY QCD with SUSY breaking generated via AMSB, ensuring that the massless spectrum matches that of QCD. They also have a QCD-like global symmetry breaking pattern after SUSY breaking is added, hence one can calculate the $\eta'$ potential and the chiral Lagrangian, as long as SUSY breaking is small compared to the scale of strong interactions. 

We find that, as expected, the resulting $\eta'$ potential has  a branched structure originating from the dynamics responsible for confinement (i.e. gluino condensation). Such branched structure cannot originate from pure instanton effects, and indeed we see that for most cases the dynamics responsible for the $\eta'$ mass is different than instantons. For a generic number $F$ of flavors we find $|N-F|$ branches, implying that the introduction of flavor qualitatively changes the confining potential. For $F<N-1$ the flavor effect is simply the breaking of the gauge group to $SU(N-F)$ via squark VEVs, while for $F>N+1$ one has a whole new $SU(F-N)$ dual gauge group, which will provide the gaugino condensates. For the special cases of $F=N-1,N,N+1$ we find a single branch for the confining potential, consistent with the entire potential being generated by a single instanton. We also find that $m_{\eta'}\to 0$ for large $N$ as long as the number of flavors is held fixed, in agreement with the expectation that the anomaly vanishes in this limit. However for $F\propto N$ the $\eta'$ mass does not vanish in the large $N$ limit, in accordance with the fact that the anomaly also does not vanish. 

We have also provided a review of the standard lore about the $\eta'$ in ordinary QCD and the dynamical origin of the axion potential, with emphasis on the large $N$ expansion. As part of this review we also presented a simple derivation of the  axion mass for an arbitrary number of flavors. In most cases the axion mass is not generated by instanton effects, hence trying to derive the axion mass formula by closing up legs on 't Hooft operators is not very useful.  
\section{Acknowledgements} 
 We thank Adi Armoni, Michael Dine, Hyung Do Kim, Zohar Komargodsky, Hitoshi Murayama, Ofri Telem and Lian-Tao Wang for useful comments and discussions. CC thanks the Tata Institute for Fundamental Research in Mumbai and the Kavli IPMU in Tokyo for its hospitality while this paper was prepared. RTD acknowledges the hospitality of the Ecole de Physique des Houches while part of this paper was completed. CC and MR are supported in part by the NSF grant PHY-2014071. MR is also supported by a Feodor–Lynen Research Fellowship awarded by the Humboldt Foundation. CC is also supported in part by a Simons Foundation Sabbatical Fellowship. CC and EK are funded in part by the US-Israeli BSF grant 2016153. 

\appendix
\section*{Appendix}
\section{$N$-dependence of the holomorphic scale}\label{app:holomScale}
The two-loop expression for the RGE invariant scale $\Lambda_c$ defined with the canonical coupling $g_c$ is constant in the large $N$ limit. This can be easily seen from the explicit expression
\begin{equation}
\Lambda_c = \mu \left( \frac{b_0 g_c^2(\mu)}{8\pi^2}\right)^{-b_1/(2b_0^2)} \exp\left( -\frac{8\pi^2}{b_0 g_c^2 (\mu)} \right)\,,
\end{equation}
where $g_c^2(\mu)$ always appears with $b_0$. For SQCD with $F$ flavors the one-loop and two-loop beta function coefficients are given by $b_0 = 3N -F$ and $b_1 = 6N^2 - 2NF-4F(N^2-1)/(2N)$.

However, the scale which appears in the superpotential is the holomorphic scale $\Lambda$ which contains the holomorphic coupling constant $\tau$ whose RGE evolution is one-loop exact, 
\begin{equation}
\Lambda = \mu e^{ \frac{2\pi i \tau(\mu)}{b_0}}\,,
\end{equation}
where $\mu$ is a holomorphic scale. In contrast to $\Lambda_c$ the holomorphic scale has a non-trivial $N$ dependence~\cite{Randall:1998ra,Armoni:2003yv,Dine:2016sgq}. The relation between the canonical and holomorphic coupling can be derived from the non-trivial Jacobian arising from the transformation to canonical gauge fields~\cite{Arkani-Hamed:1997qui,Dine:2011gd}. This gives the Shifman-Vainshtein formula~\cite{Shifman:1986zi} 
\begin{equation}
2\pi\text{Im} (\tau) = \frac{8\pi^2}{g_c^2}  + 2 T(Ad) \log g_c + \sum_i T(i) \log Z_i\,,
\end{equation}
where the sum runs over the matter fields and $T(i)$ are the Dynkin indices of representation $i$. For $SU(N)$ the Dynkin indices for the adjoint and fundamental are $T(Ad)=N$ and $T(fund)=1/2$. With $Z_i(M,\mu) = C(M) g(\mu)^{-2(N^2-1)/(b_0 N)}$,\footnote{This has been obtained from $\gamma_i \equiv (\mu d/d\mu ) \log Z_i (M,\mu) = \tfrac{g_c^2}{8\pi^2} \tfrac{N^2-1}{N} + \mathcal{O}(g^4)$ taken from~\cite{Martin:1993zk}.}  where $M$ is the cutoff and $C(M)$ a cutoff dependent constant,
we obtain
\begin{equation}
|\Lambda | = \tilde{C}(M) g_c(\mu)^{-b_1/b_0^2} \mu\exp\left( -\frac{8\pi^2}{b_0 g_c^2 (\mu)} \right)  = \tilde{C}(M) \left( \frac{b_0}{8\pi^2} \right)^{b_1/(2b_0^2)} \Lambda_c \,,
\end{equation}
%
where $\tilde{C}(M)$ is a function of $C(M)$. Since the theory is asymptotically free we can set $C(M)=\tilde{C}(M)=1$ if we take $M$ large enough.
This implies that for $N \gg F$ the holomorphic coupling scales as $|\Lambda |\propto N^{1/3} \Lambda_c$ while for $F\sim N \gg 1$ it scales as $|\Lambda | \propto N^{1/4} \Lambda_c$. In the main text we also use $\Lambda_{\rm phys}$ which absorbs the numeric prefactor in front of $\Lambda_c$ into the definition of the scale but leaves the $N$ dependence explicit. For $F=0$ this implies $|\Lambda| = N^{1/3} |\Lambda_{\rm phys}|$. 
%
\section{Axion mass for $F$ flavors}\label{app:axionMass}
%
In this appendix we outline the calculation of the leading order expression for the axion mass in the presence of $F$ massive quark flavors. The starting point is Eq.~\eqref{eq:axionpot}, i.e. the potential for the axion and neutral GBs
\begin{equation}
    V_{axion}= -2\alpha \Lambda^2 f_\pi^2 \sum_{i=1}^F \frac{m_i}{\Lambda} \left[ \cos\left(\frac{\bar{\theta} + n\, a/f_a}{F} +\sum_{j=1}^{F-1} t_i^j \frac{\pi^j}{f_\pi}\right)\right]\,,
\end{equation}
where we have reintroduced the pion and axion decay constant. Using the following parameterization for the Cartan generators
\begin{equation}
    t^j= b^j\, {\rm diag}(\underbrace{1,\ldots ,1}_j , -j,0,\ldots ,0)\,,\quad\text{with}\quad b^j = \sqrt{\frac{2}{j(j+1)}}\,,
\end{equation}
the potential can be expressed as
\begin{equation}\label{eq:axionPotPion}
    V_{axion}= -2\alpha \Lambda^2 f_\pi^2 \sum_{i=1}^F \frac{m_i}{\Lambda} \left[ \cos\left(\frac{\bar{\theta} + n\, a/f_a}{F} +\sum_{j=i}^{F-1} b^j \frac{\pi^j}{f_\pi} - (i-1) b^{i-1} \frac{\pi^{i-1}}{f_\pi}\right)\right]\,.
\end{equation}
As was already mentioned in Section~\ref{sec:axionMass} the potential is minimized for $n\, a/f_a = -\bar{\theta}$ and $\pi^j=0$. Thus we can expand the cosines to describe the quantum fluctuations around this minimum. Since $f_a \gg f_\pi$ the pions are substantially heavier than the axion and can be integrated out. As we are only interested in the axion mass it is sufficient to expand the potential to quadratic order in the fields. Thus the equation of motion for the pions is a linear equation
\be
\left[- l m_{l+1}\delta_{i,l+1} +\sum_{i\leq l}m_i\right]\left(-(i-1)b^{i-1}\pi^{i-1}+\sum_{k=i}^{F-1}\pi^k b^k\right)=-\frac{a n}{F}\frac{f_\pi}{f_a}\left[- l m_{l+1} +\sum_{i\leq l}m_i\right]\, , \nn \\
\ee
or in matrix notation
\begin{equation}
    A \begin{pmatrix}
        \pi^1/f_\pi\\ \pi^2/f_\pi \\ \vdots\\ \pi^{F-1}/f_\pi
    \end{pmatrix} = \frac{n\, a/ f_a}{F} \begin{pmatrix}
        m_2 - m_1\\ 2m_3 - m_2 - m_1\\ \vdots \\ (F-1) m_{F} - m_{F-1}-\ldots -m_1
    \end{pmatrix}\,,
\end{equation}
with the matrix $A$ given by
\begin{equation}
A=
    \begin{pmatrix}
        (1^2 m_2 +m_1) b^1  &   -(m_2 - m_1) b^2    &   \cdots  &   -(m_2 - m_1) b^{F-1}\\
        -(m_2 -m_1) b^1     &   (2^2 m_3 +m_2 +m_1) b^2 &   \cdots  & -(2 m_3 -m_2 -m_1) b^{F-1}\\
        \vdots  &   \vdots  &   &   \vdots\\
         -(m_2 -m_1) b^1    &   -(2m_3 - m_2 -m_1)b^2   &   \cdots  &   ((F-1)^2 m_F + m_{F-1}+\ldots +m_1)b^{F-1}
    \end{pmatrix}\,.
\end{equation}
For a given $F$ the equation of motion can be solved analytically. Substituting the solution into Eq.~\eqref{eq:axionPotPion} and expanding to quadratic order one finds the axion mass
\begin{equation}
   f_a^2 m_a^2 = 2\alpha \Lambda n^2 f_\pi^2
    \frac{\prod_{i=1}^F m_i}{m_1 \cdot m_2 \ldots \cdot m_{F-1} + {\rm permutations}}\,.
\end{equation}


\bibliographystyle{JHEP}
\bibliography{draft_02_MR}


\end{document}